\documentclass[final,3p,times,11pt]{elsarticle}

\pdfoutput = 1
\usepackage{bm}
\usepackage{color}
\usepackage{graphicx}
\usepackage{amssymb,amsmath,amsfonts,amsbsy}
\usepackage{cancel}

{\newcommand{\lsim}{\mbox{\raisebox{-.6ex}{~$\stackrel{<}{\sim}$~}}}
{\newcommand{\gsim}{\mbox{\raisebox{-.6ex}{~$\stackrel{>}{\sim}$~}}} 
{
\newcommand{\be}{\begin{equation}}
\newcommand{\ee}{\end{equation}}
\newcommand{\bea}{\begin{eqnarray}}
\newcommand{\eea}{\end{eqnarray}}

\newcommand{\bean}{\begin{eqnarray*}}
\newcommand{\eean}{\end{eqnarray*}}

\newcommand{\calL}{{\cal L}}

\newcommand{\calO}{{\cal O}}
\newcommand{\calP}{{\cal P}}

\newcommand{\calR}{{\cal R}}

\def\GeV{{\rm \ GeV}}

\DeclareMathAlphabet{\mathpzc}{OT1}{pzc}{m}{it}

\journal{Nuclear Physics B}

\begin{document}

\begin{frontmatter}

\begin{flushright}
{\large \tt 
TTK-12-20
\\
CERN-PH-TH/2012-140}
\end{flushright}

\title{Unifying darko-lepto-genesis with scalar triplet inflation}

\author[RWTH]{Chiara Arina}\ead{chiara.arina@physik.rwth-aachen.de}
\author[Cern]{Jinn-Ouk Gong}\ead{jinn-ouk.gong@cern.ch}
\author[IIT]{Narendra Sahu}\ead{nsahu@iith.ac.in}
\address[RWTH]{Institut f\"ur Theoretische Teilchenphysik und Kosmologie, RWTH Aachen, 52056 Aachen, Germany}
\address[Cern]{Theory Division, CERN, CH-1211 Gen\`eve 23, Switzerland}
\address[IIT]{Department of Physics, IIT Hyderabad, 
Yeddumailaram 502 205, Andhra Pradesh, India}

\begin{abstract}

We present a scalar triplet extension of the standard model to unify the origin of inflation with neutrino mass, asymmetric dark matter and leptogenesis. In presence of non-minimal couplings to gravity the scalar triplet, mixed with the standard model Higgs, plays the role of inflaton in the early Universe, while its decay to SM Higgs, lepton and dark matter simultaneously generate an asymmetry in the visible and dark matter sectors. On the other hand, in the low energy effective theory the induced vacuum expectation value of the triplet gives sub-eV Majorana masses to active neutrinos. We investigate the model parameter space leading to successful inflation as well as the observed dark matter to baryon abundance. Assuming the standard model like Higgs mass to be at 125-126 GeV, we found that 
the mass scale of the scalar triplet to be $\lsim O(10^9)$ GeV and its trilinear coupling to doublet Higgs is 
$\lsim 0.09$ so that it not only evades the possibility of having a metastable vacuum in the standard model, but also lead to a rich phenomenological consequences as stated above. Moreover, we found that the scalar triplet 
%. in order to prevent vacuum instability at a scale $\lsim O(10^8)$ GeV. It is found that 
inflation strongly constrains the quartic couplings, while allowing for a wide range of Yukawa couplings which generate the CP asymmetries in the visible and dark matter sectors.

\end{abstract}


\end{frontmatter}

\section{Introduction}
\label{sec:intro}

A widely accepted theory of the early Universe supposes that there has been a period of cosmic inflation~\cite{Guth:1980zm,Linde:1981mu,Albrecht:1982wi} which not only explains the drawbacks of standard cosmology, but also provides seed for the temperature anisotropy in the cosmic microwave background~\cite{Lyth:1998xn,Mukhanov:2005sc,Martin:2007bw,Lyth:2009zz,Mazumdar:2010sa}. Finding a particle physics model for the inflaton is a non-trivial task however. In the standard model (SM) of particle physics, the only scalar field is the $SU(2)$ doublet Higgs , whose quartic coupling $\lambda_H$ is not a free parameter once its mass is fixed. Hence a model of chaotic inflation is not possible within the framework of SM. However, by adding one more coupling $\xi_H$ between the Higgs and gravity~\cite{Spokoiny:1984bd,Accetta:1985du,Salopek:1988qh}, the potential could be made flat enough for producing approximately 60 $e$-folds of inflation. Indeed there is a plateau for value of the field $h \gg M_{\rm pl}/\sqrt{\xi_H}$, where $M_{\rm pl}$ is the reduced Planck mass. The phenomenological inflationary constraints are met when $\lambda_H/\xi_H^2$ matches the amplitude of density perturbations. For instance with a quartic coupling of $\mathcal{O}(0.1)$ the non-minimal coupling to gravity $\xi_H$ is bounded to be $\mathcal{O}(10^4)$, and hence inflation takes place at the unitarity scale $M_{\rm pl}/\xi_H \simeq 10^{14}$ GeV~\cite{Burgess:2009ea,Barbon:2009ya,Hertzberg:2010dc,Lerner:2010mq}. This is the so-called Higgs inflation~\cite{Bezrukov:2007ep,Bezrukov:2008ut,Bezrukov:2010jz}. However, the indication of SM like Higgs at 125-126 GeV~\cite{ATLAS,CMS} lead to a metastable vacuum~\cite{EliasMiro:2011aa,ArkaniHamed:2008ym} at around $10^9$ GeV, which is much below the unitarity scale. The current uncertainties in the experimental measurements although allow 
one to extend the vacuum instability up to Planck scale, but it can only be resolved at future experiments. One of 
the possibilities to evade this issue is to widen the scalar field content of the SM. Extension of Higgs inflation 
by means of a scalar singlet or the inert doublet have been discussed in~\cite{Lerner:2009xg,Lebedev:2011aq,Gong:2012ri,Lebedev:2012zw,EliasMiro:2012ay}.

It is paramount to restore a thermal bath at the end of inflation to generate visible and dark matter (DM) observed today. At present a number of evidences suggests the existence of DM, which constitutes one quarter of the total energy budget of the Universe~\cite{Bertone:2004pz,Komatsu:2010fb}. However, hitherto a definite mechanism that gives rise to the observed relic abundance of DM is unknown. Usually it is assumed that the DM particle is in thermal equilibrium in the early Universe and freeze-out below its mass scale~\cite{Kolb:1990vq}. However, an alternative scenario to the freeze-out mechanism is that the relic abundance of DM can be accounted by an asymmetric component rather than by the symmetric one~\cite{Nussinov:1985xr, Barr:1990ca,Dodelson:1991iv,Kaplan:1991ah,Kuzmin:1996he,Fujii:2002aj,Oaknin:2003uv, Hooper:2004dc,Kitano:2004sv,Cosme:2005sb,Farrar:2005zd,Roszkowski:2006kw, McDonald:2006if,Kohri:2009yn,An:2009vq,Kaplan:2009ag,Shelton:2010ta, Davoudiasl:2010am,Haba:2010bm,Gu:2010ft,Blennow:2010qp,McDonald:2010rn, Hall:2010jx,Dutta:2010va,Falkowski:2011xh,Chun:2011cc,Cui:2011ab,Arina:2011cu, Barr:2011cz,Petraki:2011mv,Iminniyaz:2011yp,Graesser:2011wi,Buckley:2011kk, Kouvaris:2011gb,Cirelli:2011ac,vonHarling:2012yn,Davoudiasl:2012uw,Tulin:2012re, Blennow:2012de}. Since none of the particles in the SM can be a candidate of DM, one needs to explore physics beyond SM to have a particle physics candidate for DM. Apart from DM, the non-zero neutrino masses as confirmed by the oscillation data are required to be explained in a beyond SM framework. Recall that neutrinos are exactly massless within SM because of the conservation of lepton number up to all orders in perturbation theory.

Besides DM and neutrino mass, an explanation for the observed matter-antimatter asymmetry required for the big bang nucleosynthesis is still missing within the framework of SM. If the reheating temperature is less than electroweak (EW) scale then it is difficult to generate both DM and the observed baryon asymmetry~\cite{Kohri:2009ka}. On the other hand, if the reheating temperature is larger than EW scale, several mechanisms are available which can give rise to required baryon asymmetry, while leaving a large temperature window for creating DM species observed today. In the past years a lot of effort have been made to unify the mechanism giving rise to the asymmetry both in the DM and baryonic sectors~\cite{Nussinov:1985xr,Barr:1990ca,Dodelson:1991iv,Kaplan:1991ah,Kuzmin:1996he, Fujii:2002aj,Oaknin:2003uv,Hooper:2004dc,Kitano:2004sv,Cosme:2005sb,Farrar:2005zd, Roszkowski:2006kw,McDonald:2006if,Kohri:2009yn,An:2009vq,Shelton:2010ta, Davoudiasl:2010am,Haba:2010bm,Gu:2010ft,Blennow:2010qp,McDonald:2010rn,Hall:2010jx, Dutta:2010va,Falkowski:2011xh,Chun:2011cc,Cui:2011ab,Arina:2011cu,Barr:2011cz, Petraki:2011mv,vonHarling:2012yn,Davoudiasl:2012uw,Walker:2012ka,Heckman:2011sw,MarchRussell:2012hi,Frandsen:2011kt,Belyaev:2010kp}. An attempt to unify DM and baryon asymmetry via leptogenesis route has also been proposed by two of the authors in~\cite{Arina:2011cu}, where SM is extended by introducing a $SU(2)_L$ scalar triplet and a fermionic doublet dark matter candidate, stable by means of a remnant $Z_2$ flavour symmetry. The triplet is taken to be at high scale such that its out-of-equilibrium decay can produce asymmetric DM as well as visible matter through leptogenesis mechanism~\cite{Ma:1998dx,Hambye:2000ui}. Moreover, in the low energy effective theory the induced vacuum expectation value (vev) of the scalar triplet could give rise sub-eV Majorana masses to the active neutrinos. Thus a triple unification of asymmetric DM, baryon asymmetry and neutrino masses in a minimal extension of the SM is achieved.

In this article, we realize primordial inflation in the presence of non-minimal coupling $\xi_\Delta$ to gravity in a scalar triplet ($\Delta$) extension of the SM and study the consequent low energy phenomenology. An early attempt of triplet inflation has been discussed in~\cite{Chen:2010uc} within the framework of chaotic inflation, where the quartic coupling of the triplet is supposed to be negligibly small (less than $10^{-13}$) and the dominant term in the scalar potential is the triplet mass, around $10^{13}$ GeV. In presence of the non-minimal coupling of the scalar triplet to gravity the mass scale of the triplet can be much below than $10^{13}$ GeV without fine tuning the quartic coupling. We take the mass scale of triplet to be around $10^8-10^9$ GeV such that it not only give neutrino masses, dark matter abundance and baryon asymmetry, but also evade the possiblity of having a metastable vacuum in the SM~\cite{EliasMiro:2011aa,ArkaniHamed:2008ym}. In presence of non-minimal couplings $\xi_\Delta$ and $\xi_H$ to gravity the scalar triplet, together with the SM Higgs field, behaves as inflaton. From this multi-field inflationary scenario a single field model can be retrieved as we demonstrate below. We show that once the heavy mode is settled down at the minimum, the scalar potential is positive definite only if the mass term and the lepton number violating term ($\mu_H \Delta^\dagger HH$) are negligible. However, the inflaton can be an admixture of both triplet and SM Higgs moduli or a pure state. We demonstrate in detail how these three cases give rise to different constraints on the model parameter space. Subsequently, we explain how the decay of scalar triplet~\cite{Arina:2011cu} can generate an asymmetric dark matter and visible matter observed today.

The article is organized as follows. In section~\ref{sec:pot} we briefly underline the main features of the model, which has been introduced in~\cite{Arina:2011cu} and point out new constraints in the parameter space. We then describe the inflationary picture in section~\ref{sec:inflation}, where we work out the slow-roll predictions for single field inflation after having discussions regarding the numerical and analytical estimates of all the terms in the scalar potential. The generation of the asymmetries in the dark and visible sectors are discussed in section~\ref{sec:ADML}. The ensuing section~\ref{sec:RGEs} details the renormalization group (RG) equations accounting for the additional field content with respect to the SM ones. Our results are presented in section~\ref{sec:res} and we conclude in section~\ref{sec:concl}. We recall in~\ref{appA} the main Boltzmann equations for the production of the asymmetries in both baryonic and DM sectors.

\section{Scalar Triplet as the Origin of Inflation and Darko-Lepto-genesis}
\label{sec:pot}

We extend SM by introducing a scalar triplet $\Delta (3,2)$, where the quantum numbers in the parenthesis are the charge under the gauge group $SU(2)_L \times U(1)_Y$. Since the hypercharge of $\Delta$ is 2, it can have bilinear coupling to the Higgs doublet $H$. As a result the scalar potential involving $\Delta$ and $H$ can be given as follows:
\begin{equation}\label{eq:ScalarPotential}
V_J(\Delta, H) =  M_\Delta^2 \Delta^\dagger \Delta + \frac{\lambda_\Delta}{2} (\Delta^\dagger \Delta)^2 -  M_H^2 H^\dagger H + \frac{\lambda_H}{2} (H^\dagger H)^2 + \lambda_{\Delta H} H^{\dagger} H 
\Delta^\dagger \Delta + \frac{1}{\sqrt{2}} \left[ \mu_H \Delta^\dagger H H + {\rm h.c.} \right]  \,,
\end{equation}
where the index $J$ stands for the Jordan frame, as will be explained in the next section~\ref{sec:inflation}. The $2\times 2$ representation of the scalar triplet is 
 \begin{equation}
\Delta = \begin{pmatrix} 
\Delta^+/\sqrt{2} & \Delta^{++}
\cr
\Delta^0 &  -\Delta^+/\sqrt{2} 
\end{pmatrix}\,.
\end{equation}

In the fermion sector we introduce a vector-like doublet $\psi \equiv (\psi_{\rm DM}, \psi_-)$ with hypercharge $Y=-1$~\cite{Arina:2011cu}. As a result the bilinear couplings of $\Delta$ to the lepton doublets $L$, $\psi$ and $H$ are given as follows:
\begin{equation}\label{eq:Lag-DM}
 -\mathcal{L}  \supset  \overline{\psi} i\gamma^\mu \mathcal{D}_\mu \psi + M_D \overline{\psi} \psi  +  \frac{1}{\sqrt{2}} \left[ f_H \Delta^\dagger H H + f_L \Delta L L + f_\psi \Delta \psi \psi + {\rm h.c.} \right]\,,
\end{equation}
where $f_H=\mu_H/M_\Delta$. The covariant derivative $\mathcal{D}_\mu$ is defined as 
\begin{equation}
\mathcal{D}_\mu = \partial_\mu + i \sqrt{\frac{3}{5}}g_1 B_\mu + i g_2 t W_\mu\,,
\end{equation} 
where $t$ represents the Pauli spin matrices. For the hypercharge coupling we have used the grand unified theory (GUT) charge normalisation: $3\left( g_1^{\rm GUT}\right)^2/5 = \left( g_1^{\rm SM}\right)^2$.

From (\ref{eq:ScalarPotential}) and (\ref{eq:Lag-DM}) we notice that:
\begin{enumerate}
\item The bilinear coupling of $\Delta$ to the Higgs and lepton doublets jointly violate lepton number by two units. Moreover, the couplings are complex and hence can accommodate a net CP violation. As a result the out-of-equilibrium decay of $\Delta$ to $LL$ and $HH$ in the early Universe can give rise to the observed matter-antimatter asymmetry via leptogenesis route~\cite{Ma:1998dx,Hambye:2000ui}.

\item The Lagrangian is invariant under a remnant $Z_2$ symmetry, with $\psi$ being odd while all the other fields even. This ensures the stability of $\psi_{\rm DM}$, the neutral component of $\psi$, which can be a candidate of dark matter. Hereafter $\psi_{\rm DM}$ is the inert fermion doublet DM~\cite{Arina:2011cu}. Since the bilinear coupling of $\Delta$ to $\psi \psi$ is in general complex, it can accommodate a net CP violation. Therefore, the out-of-equilibrium decay of $\Delta \to \psi \psi$ in the early Universe can generate an asymmetry in DM sector in a similar way the lepton asymmetry is generated via the decay $\Delta \to L L$ and $\Delta \to HH$.

In the effective theory the bilinear coupling of $\Delta$ to $HH$ and $\psi\psi$ generates a dimension-five operator ${\cal O}_5=\psi\psi H H$ suppressed by the mass scale of $\Delta$. This is an equivalent type-II seesaw for Majorana mass of DM. Below EW phase transition this operator generates small Majorana mass for $ \psi_{\rm DM}$ as given by
\begin{equation}
m= \sqrt{2} f_\psi \langle \Delta \rangle = f_H f_\psi \frac{-v^2}{M_\Delta}\,,
\end{equation}
where $v=\langle H \rangle $ is the vev of the SM Higgs. Since $\psi_{\rm DM}$ is a vector-like Dirac fermion, it can be expressed as a sum of two Majorana fermions, i.e. $\psi_{\rm DM} =\left( \psi_{\rm DM} \right)_L + \left( \psi_{\rm DM} \right)_R$.  Therefore, in a flavour basis $((\psi_{\rm DM})_L, (\psi_{\rm DM})^c_R)$, the mass matrix of DM is given by
\begin{equation}
\mathcal{M} = \begin{pmatrix} 
M_D & m/2 
\cr 
m/2 & M_D 
\end{pmatrix}
\,.
\end{equation}
Diagonalising the above mass matrix we get two mass eigenstates $(\psi_{\rm DM})_1$ and $(\psi_{\rm DM})_2$ with masses $M_D + m/2$ and $M_D-m/2$. The mass splitting $\delta \sim m $ between the two states is required to be ${\cal O}(100)$ keV in order to explain the high precision annual modulation signal at DAMA~\cite{Bernabei:2010mq,TuckerSmith:2001hy,Arina:2009um,Arina:2011cu} while the null result at Xenon100~\cite{Aprile:2011ts}. This implies a lower bound on $f_\psi$ to be
\begin{equation}
f_\psi = \frac{m}{\sqrt{2} \langle \Delta \rangle } \gsim 10^{-4} \,,
\end{equation}
where we have assumed  $\langle \Delta \rangle \lsim {\cal O}(1) \GeV$ as required by the $\rho$ parameter of SM.

\item In the effective low energy theory the bilinear coupling of $\Delta$ to lepton and Higgs doublets also generate a dimension-five operator ${\cal O}_5 = LLHH$, suppressed by the mass scale of $\Delta$, for neutrino masses. When $H$ acquires a vev, this operator then induces sub-eV Majorana masses to active neutrinos given by:
\begin{equation}
M_\nu =  \sqrt{2} f_L \langle \Delta \rangle = f_L f_H \frac{-v^2}{M_\Delta}\,.
\end{equation}
For $M_\Delta \gg v$, we can easily obtain sub-eV masses of active neutrinos for a wide range of values of the couplings $f_L$ and $f_H$. For example, taking $f_L$ and $f_H$ to be order unity we need $M_\Delta \sim 10^{12}$ GeV to get sub-eV neutrino masses. For lighter $\Delta$ one can get neutrino masses in the ball park of oscillation data by taking smaller values of $f_L$, yet maintaining vev of $\Delta$ to be less than ${\cal O}(1)$ GeV. An advantage for smaller values of $f_L$ is that we can easily explain the required ratio:
\begin{equation}
R \equiv \frac{M_\nu}{m} =\frac{f_L}{f_\psi} \approx {\cal O}(10^{-5})\,.
\end{equation}
Thus for $f_\psi \gsim 10^{-4}$, we expect $f_L\gsim 10^{-9}$.

\item In the presence of the non-minimal couplings of $\Delta$ and $H$ to gravity, the scalar potential (\ref{eq:ScalarPotential}) can give rise to inflation in the early Universe~\cite{Bezrukov:2007ep,Bezrukov:2008ut,Bezrukov:2010jz}. The scale of inflation at which the power spectrum is normalized (see later section) is $\left[ V(\Delta,H)/\epsilon\right]^{1/4} \simeq 10^{16}$ GeV, which is much below the Planck scale. At the end of inflation, the Universe becomes radiation dominated, during which the interactions of $\Delta$ as given in (\ref{eq:Lag-DM}) generate asymmetries in visible and DM sectors. 
\end{enumerate}

\section{Scalar Triplet -- Higgs Inflation}
\label{sec:inflation}

\subsection{Action in the Einstein frame}

The model for the scalar fields has been defined in the previous section. The scope of this section is to work out the action for inflation. The physical fields are defined in the Jordan frame denoted by an index $J$. We introduce for both scalar components non-minimal couplings to the Ricci scalar $R$. Hence the action in the Jordan frame is:
\begin{equation}
S_{J}  =  \int {\rm d}^4 x \,\sqrt{-g}\,  \left[ \frac{R}{2}
 +   \left(\xi_H H^{\dagger} H + \xi_{\Delta} \Delta^{\dagger}\Delta + c.c.\right) \ R 
 -  |\mathcal{D}_{\mu} H|^2 - |\mathcal{D}_{\mu} \Delta|^2 - V_J(H,\Delta)\right] \,,
\label{eq:Sj}
\end{equation}
with the reduced Planck mass set to unity, i.e. $M^2_{\rm pl} = m^2_{\rm pl}/(8 \pi) = 1$.

In the Jordan frame the couplings $\xi_i$ make the gravitational interactions non-standard. It is therefore convenient to perform a conformal transformation into the Einstein frame, for which we put no index, to retrieve the standard form of the Einstein equations as far as gravity concern, but at the expense of having non-standard kinetic terms for the scalar fields. A conformal transformation preserves the causal structure of space-time in both frames and is given by a smooth and strictly positive function of the fields:
\begin{equation}
\Omega^2 = 1 + 2 \xi_\Delta |\Delta|^2
+ 2 \xi_H |H|^2
\,.
\label{eq:ct}
\end{equation}
Note that both frame are equivalent for small field values. The metric and the potential transform as:
\begin{align}
\tilde{g}_{\mu\nu} = & \, \, \Omega^2 g_{\mu\nu}^J\,,
\label{eq:g}  
\\
V(H,\Delta) = &\, \,  \frac{V_J(H,\Delta)}{\Omega^4}\,.
\label{eq:scalpote}
\end{align}
The doublet and triplet scalar fields are defined in the unitary gauge as following:
\begin{align}
 H = &\,  \frac{1}{\sqrt{2}} 
 \left(
 \begin{array}{c}
 0 \\h
 \end{array}
 \right)\,,
\\
 \Delta = &\, \frac{1}{\sqrt{2}} \left(\begin{array}{cc}0 & 0 \\ \delta e^{i \theta} & 0\end{array}\right)\,,
\end{align}
where $\delta$ and $\theta$ account for the two degrees of freedom of the triplet neutral component, defined as $\Delta^0 = \left[{\rm Re}(\Delta^0) + i \ {\rm Im}(\Delta^0)\right]/\sqrt{2}$.

Now taking the large field limit $\xi_\Delta \delta^2 + \xi_H h^2 \gg 1$ and redefining fields as:
\begin{align}
\label{eq:sffin}
\varphi = &\,  \sqrt{\frac{3}{2}} \log\left(1+\xi_\Delta\delta^2+\xi_Hh^2\right) \, ,
\\
r = & \, \frac{\delta}{h} \, ,
\end{align}
(\ref{eq:Sj}) reads:
\begin{align}\label{eq:FinS}
S  = & \int {\rm d}^4 x \,\sqrt{-\tilde{g}}\,  \left[ \frac{\widetilde{R}}{2}
 -   \frac{1}{2} \left(1+\frac{1}{6}\frac{r^2+1}{\xi_H + \xi_\Delta r^2}\right) (\partial_{\mu} \varphi)^2
 -  \frac{1}{\sqrt{6}} \frac{(\xi_H-\xi_\Delta) r}{(\xi_H + \xi_\Delta r^2)^2} (\partial_{\mu} \varphi)(\partial_{\mu} r)\right.
\nonumber\\
 & \hspace{2cm} - \left.\frac{1}{2} \frac{\xi^2_H + \xi^2_\Delta r^2}{(\xi_H + \xi_\Delta r^2)^3} (\partial_{\mu} r)^2
 -  \frac{1}{2} \frac{r^2}{\xi_H + \xi_\Delta r^2}\left( 1-e^{-2\varphi/\sqrt{6}} \right) (\partial_{\mu} \theta)^2 - V(r,\varphi,\theta) \right]\,.
\end{align}
Note that the kinetic part is highly non-trivial for all fields $\varphi$, $r$ and $\theta$. However the potential, with the field redefinition, takes the form:
\begin{align}\label{potential}
V(r,\varphi,\theta) = &\,  \frac{\lambda_H/2+\lambda_{H\Delta}r^2+\lambda_\Delta r^4/2}{4(\xi_H+\xi_\Delta r^2)^2} \left( 1-e^{-2\varphi/\sqrt{6}} \right)^2 + \frac{M_H^2+M_\Delta^2r^2}{2(\xi_H+\xi_\Delta r^2)}e^{-2\varphi/\sqrt{6}}\left( 1-e^{-2\varphi/\sqrt{6}} \right)
\nonumber\\
&+ \frac{\mu_Hr\cos\theta}{2(\xi_H+\xi_\Delta r^2)^{3/2}} e^{-\varphi/\sqrt{6}} \left( 1-e^{-2\varphi/\sqrt{6}} \right)^{3/2} \, .
\end{align}

\subsection{Scalar potential analysis}
\label{sec:potanalyt}

During inflation the mass eigenvalue of $r$ is very large as compared to the Hubble parameter~\cite{Gong:2012ri}. Therefore, $r$ is minimized at $r_0$ and we find the effective theory for the light inflatons. The action then becomes:
\begin{equation}
\frac{\calL}{\sqrt{-\tilde{g}}}  =  -\frac{1}{2} \left[ 1 + \frac{1+r_0^2}{6(\xi_H+\xi_\Delta r_0^2)} \right] (\partial_\mu\varphi)^2 - \frac{1}{2}\frac{r_0^2}{\xi_H+\xi_\Delta r_0^2} \left( 1-e^{-2\varphi/\sqrt{6}} \right) (\partial_\mu\theta)^2 - V(\varphi,\theta) \, ,\\
\end{equation}
with $V(\varphi,\theta)   =  V(r \to r_0,\varphi,\theta)$. Note that the stabilization of $r$ demands important constraints on the couplings, which will be discussed in the following section.
For a finite value of $r_0$, with
\begin{align}
\lambda_{\rm eff} = & \, \frac{\lambda_H}{2} + \lambda_{H\Delta}r_0^2 + \frac{\lambda_\Delta}{2} r_0^4 \, , 
\\
\xi_{\rm eff}  = &\,  \xi_H + \xi_\Delta r_0^2 \, ,
\end{align}
we can further approximate the kinetic sector as
\begin{equation}
\frac{\calL_{\rm kin}}{\sqrt{-\tilde{g}}} = \frac{1}{2}(\partial_\mu\varphi)^2 + \frac{1}{2} \left( 1-e^{-2\varphi/\sqrt{6}} \right) (\partial_\mu\chi)^2 \, ,
\end{equation}
where $\chi=\theta r_0/\sqrt{\xi_{\rm eff}}$.

For the potential, as can be seen from (\ref{potential}), it consists of three contributions -- quartic, quadratic and the $\mu$-terms. Since the latter two are exponentially suppressed, one may be tempted to drop them from the beginning for simplicity. However we must check explicitly if quartic term is really dominant, only after then we can make any simplification. First let us compare the quartic term with the quadratic mass term:
\begin{equation}
\frac{V_M}{V_\lambda} \sim M_\Delta^2r_0^2e^{-2\varphi/\sqrt{6}}\frac{\xi_{\rm eff}}{\lambda_{\rm eff}} \, .
\end{equation}
Here we first {\em assume} the quartic term is dominant, which normalizes the combination $\lambda_{\rm eff}/\xi_{\rm eff}^2 \sim 10^{-9}$ from the amplitude of the power spectrum (see later section). We will justify this assumption a posteriori. Then, with the typical value of $\varphi$ during inflation, say $\varphi \sim 5$, we have $e^{-2\varphi/\sqrt{6}} \sim 10^{-2}$ so that the ratio becomes
\begin{equation}\label{ratioVMVl}
\frac{V_M}{V_\lambda} \sim M_\Delta^2 \, 10^{-2}\frac{10^9}{\xi_{\rm eff}}r_0^2 \sim 10^7M_\Delta^2\frac{r_0^2}{\xi_{\rm eff}} \, .
\end{equation}
It is not difficult to set this ratio negligibly small with large enough $\xi_{\rm eff}$ and not too large $r_0$ and $M_\Delta$: for $M_\Delta \sim 10^{-6}$ ($M_\Delta \sim 10^{12}$ GeV), this ratio becomes $10^{-5}r_0^2/\xi_{\rm eff}$ which can be easily made small, and even easier if we let $M_\Delta$ smaller than $10^{-6}$. For the triplet term with $\mu_H$ we can proceed similarly, and obtain
\begin{equation}\label{ratioVmVl}
\frac{V_\mu}{V_\lambda} \sim \mu_He^{-\varphi/\sqrt{6}} \frac{1}{\lambda_{\rm eff}/\xi_{\rm eff}^2}\frac{r_0}{\xi_{\rm eff}^{3/2}} \sim 10^8\mu_H\frac{r_0}{\xi_{\rm eff}^{3/2}} \, ,
\end{equation}
which looks more stringent than $V_M/V_\lambda$ and there indeed is a tension: with large enough $r_0$ and $\mu_H$ and not too large $\xi_{\rm eff}$ this ratio may be close to 1 and we should not neglect $V_\mu$. However there is another constraint that the potential be positive everywhere. For simplicity, let us neglect $V_M$ which can be made easily negligible, then the potential is
\begin{equation}
V \sim  10^{-10} \left( 1 - e^{-2\varphi/\sqrt{6}} \right)^2 + \frac{r_0}{2\xi_{\rm eff}^{3/2}}\mu_H \cos\theta e^{-\varphi/\sqrt{6}} \left( 1 - e^{-2\varphi/\sqrt{6}} \right)^{3/2} \, ,
\end{equation}
which should be positive definite. This gives
\begin{equation}
\mu_H\frac{r_0}{\xi_{\rm eff}^{3/2}} \lesssim 10^{-10} e^{\varphi/\sqrt{6}} \left( 1 - e^{-2\varphi/\sqrt{6}} \right)^{1/2} \, .
\end{equation}
We can easily note that $e^{\varphi/\sqrt{6}} \left( 1-e^{-2\varphi/\sqrt{6}} \right)^{1/2}$ is a mildly increasing function of $\varphi$ with the values 1.12364 at $\varphi=1$ and 7.63495 at $\varphi=5$. Thus, to guarantee the positivity of the potential until the end of inflation where $\varphi_e \sim 1$ provided that $V_\lambda$ is dominant, we should demand
\begin{equation}\label{muHbound}
\mu_H\frac{r_0}{\xi_{\rm eff}^{3/2}} \lesssim 10^{-10} \, ,
\end{equation}
which in turn gives, combined with (\ref{ratioVmVl}),
\begin{equation}
\frac{V_\mu}{V_\lambda} \sim 10^8\mu_H\frac{r_0}{\xi_{\rm eff}^{3/2}} \lesssim 10^{-2} \, .
\end{equation}
That is, the positivity of the potential demands that the quartic term be dominant, with the fraction of the triplet term contribution at most $\calO(1)$ percent. Further, returning back to (\ref{ratioVMVl}), using (\ref{muHbound}) we find
\begin{equation}
\frac{V_M}{V_\lambda} \lesssim 
10^{-13} \left( \frac{M_\Delta}{\mu_H} \right)^2\xi_{\rm eff}^2 \, .
\end{equation}
Thus, for $M_\Delta \sim \mu_H$, $V_M$ remains indeed negligible compared with $V_\lambda$ unless $\xi_{\rm eff}$ is very large. However too large $\xi_{\rm eff}$ will pull down the unitarity scale further, greatly harming the validity of the effective theory: if $\xi_{\rm eff} \sim 10^6$, $V_M$ may compete with $V_\lambda$ up to $\calO(10)$ percent, but the unitarity scale $\mu_U \sim \xi_{\rm eff}^{-1} \sim 10^{-6}$ may well be saturated near $M_\Delta \sim \mu_H \sim 10^{-6}$ and the low energy approximation cannot be trusted. So not too large $\xi_{\rm eff}$ guarantees negligible contribution of $V_M$. All these a posteriori justify our assumption at the beginning that the potential is dominated by the quartic term so that $\lambda_\text{eff}/\xi_\text{eff}^2 \sim 10^{-9}$.

This estimate gives us the idea that the contributions of $\chi$ to the observable quantities are not significant. To check this, we first compute numerically the change in the number of $e$-folds $N$ as follows.
We compute $N(\varphi_\star,\chi_\star)$ from the moment $\star$, when the scale of our interest exits the horizon, to $e$, the end of inflation, with a given set of initial conditions of $\varphi_\star$ and $\chi_\star$. Then we repeats with slightly different initial conditions to find $N(\varphi_\star+\Delta\varphi,\chi_\star)$ and $N(\varphi_\star,\chi_\star+\Delta\chi)$. Then, we find $\delta{N}$ according to the change in the initial field values. In table~\ref{tab:numdn} we show $\partial{N}/\partial\varphi_\star$ and $\partial{N}/\partial\chi_\star$ for several values of $\mu_H$, $\xi_{\rm eff}$ and $r_0$ (note that the amplitude of the power spectrum fixes $\lambda_{\rm eff}$ for a given $\xi_{\rm eff}$). Single field analytic estimate (\ref{singledeltaN}) gives $\partial{N}/\partial\varphi_\star=35.6967$ (54.0031) for $\varphi_\star=5$ (5.5).

\begin{table}[t!]
\caption{Numerical estimates of the contributions of the fields $\varphi$ and $\chi$ to the total number of $e$-folds $N_0$. For the first case the bound (\ref{muHbound}) is saturated, while the rest examples satisfy it trivially.}
\label{tab:numdn}
\begin{center}
\begin{tabular}{c||c|c|c}
 & $N_0$ & $\partial{N}/\partial\varphi_\star$ & $\partial{N}/\partial\chi_\star$
\\\hline\hline
$r_0=1$, $\xi_{\rm eff}=10^4$, $\mu_H=10^{-7}$, $\varphi_\star=5$, $\chi_\star=10^{-3}$ & 42.0850 & 35.9478 & $-2.98106$
\\
the same as above but $\varphi_\star=5.5$ & 64.3191 & 54.2294 & $-5.24260$
\\
the same as above but $\chi_\star = 10^{-3.5}$ & 42.0884 & 35.9508 & $-8.89659$
\\
\hline
$r_0=10$, $\xi_{\rm eff}=10^3$, $\mu_H=10^{-9}$, $\varphi_\star=5$, $\chi_\star=10^{-3}$ & 42.1880 & 36.0828 & $-1.07484 \times 10^{-3}$
\\
the same as above but $\varphi_\star=5.5$ & 64.5161 & 54.4816 & $-1.98455 \times 10^{-3}$
\\
the same as above but $\chi_\star=10^{-3.5}$ & 42.1880 & 36.0828 & $-3.39071 \times 10^{-4}$
\\
\hline
$r_0=10^2$, $\xi_{\rm eff}=50$, $\mu_H=10^{-11}$, $\varphi_\star=5$, $\chi_\star=10^{-3}$ & 42.1785 & 36.0711 & $-4.19220 \times 10^{-7}$
\\
the same as above but $\varphi_\star=5.5$ & 64.4986 & 54.4600 & $-9.66338 \times 10^{-7}$
\\
the same as above but $\chi_\star=10^{-3.5}$ & 42.1785 & 36.0711 & $-3.55271 \times 10^{-8}$
\end{tabular}
\end{center}
\end{table}

\subsection{Constraints on the scalar potential}
\label{sec:consPot}

From now on, as discussed in the previous section, we only consider single field case where $\mu_H$ term does not contribute and only $\varphi$ drives inflation. However as mentioned at the beginning, we {\em assume} that $r$ is stabilized already. For this to happen, we need to study in detail this stabilization which gives constraints on the couplings. These constraints do affect low energy phenomenology by incorporating RG equations, even $\mu_H$ which does not participate in the inflationary dynamics but whose RG equation does include quartic couplings.

We first have to ensure that the potential, quartic terms alone, is positive definite everywhere. This is necessary because we may not have to ensure $\lambda_{H\Delta}>0$. Indeed, we must have $\lambda_H>0$, $\lambda_\Delta>0$ and
\begin{equation}
\lambda_{H\Delta} + \sqrt{\lambda_H\lambda_\Delta} > 0 \, ,
\end{equation}
for positive potential.

Coming back to the definition of the potential in terms of $r$ and $\varphi$, as the mass eigenvalue for $r$ is very large compared to $H$ (see Appendix~B of~\cite{Gong:2012ri} and~\cite{Achucarro:2010jv,Achucarro:2010da,Achucarro:2012sm}) we assume $r$ is stabilized at $r=r_0$ throughout the whole process of our interest. The different minima in which the heavy field $r$ quickly sets in, are found minimizing the potential part independent of $\varphi$: 
\begin{equation}\label{eq:vrpot}
 V_{\varphi\text{-indep}} = 
\frac{\lambda_H/2 + \lambda_\Delta/2 r^4 + \lambda_{H\Delta} r^2}{4 (\xi_H + \xi_\Delta r^2)^2}\,.
\end{equation}
The minima are listed below together with the corresponding minimum energy and constraints for vacuum stability. At $r=0$ and $r=\infty$ inflation is driven by pure Higgs ($r=0$) or pure triplet ($r=\infty$). At the finite minimum, inflation is driven by an admixture of both fields.

\begin{enumerate}

\item $r^2=(\lambda_{H\Delta} \xi_H- \lambda_H \xi_\Delta)/(\lambda_{H\Delta}\xi_\Delta-\lambda_\Delta \xi_H)$: Then, $V_{\varphi\text{-indep}}$ becomes a constant, i.e. vacuum energy, of the value:
\begin{equation}
V_{\varphi\text{-indep}} \equiv V_0^{\rm (mixed)} =\frac{\lambda_\Delta \lambda_H - \lambda_{H\Delta}^2}{8\,  (\lambda_\Delta\, \xi^2_H + \lambda_H\, \xi^2_\Delta -2\lambda_{H\Delta} \xi_\Delta \xi_H)} \,.
\end{equation}
We demand that $V_0^{\rm (mixed)}>0$ and $dV^2/dr^2|_{r^2=r_0^2}>0$. Then, we must satisfy the conditions
\begin{align}
\label{eq:constMIx1}
\lambda_H\lambda_\Delta-\lambda_{H\Delta}^2 > & \, 0 \, ,
\\
\label{eq:constMIx2}
\xi_H\lambda_{H\Delta}-\xi_\Delta\lambda_H  < & \, 0 \, ,
\\
\label{eq:constMIx3}
\xi_\Delta\lambda_{H\Delta}-\xi_H\lambda_\Delta  < & \, 0 \, .
\end{align}
Note that the first condition is also equivalent to demanding that the numerator of (\ref{eq:vrpot}), which is essentially a quadratic equation of $r^2$, is always positive, i.e. the equation has no solution of $r^2=0$.

\item $r^2 \to 0$: In this case $\delta \to 0$ so this corresponds to pure Higgs inflation, i.e. the Higgs moduli alone drives inflation. $V_{\varphi\text{-indep}}$ becomes a constant, i.e. vacuum energy, of the value
\begin{equation}
V_{\varphi\text{-indep}} \equiv V_0^{(H)} = \frac{\lambda_H}{8\xi_H^2} \,.
\end{equation}
In this case $d^2V/dr^2|_{r^2=0}>0$ gives 
\begin{align}
\label{pureHiggs_condition1}
\xi_H\lambda_{H\Delta}-\xi_\Delta\lambda_H > & 0 \, ,
\\
\label{pureHiggs_condition2}
\xi_\Delta\lambda_{H\Delta}-\xi_H\lambda_\Delta < & 0 \, .
\end{align}

\item $r^2 \to \infty$: In this case $h \to 0$ so this corresponds to pure triplet inflation (in this case the triplet moduli alone drives inflation) with:
\begin{equation}
V_{\varphi\text{-indep}} \equiv V_0^{(\Delta)} = \frac{\lambda_\Delta}{8\xi_\Delta^2} \, .
\end{equation}
In this case $d^2V/dr^2|_{r^2=\infty}>0$ gives 
\begin{align}
\label{puretriplet_condition1}
\xi_H\lambda_{H\Delta}-\xi_\Delta\lambda_H < & 0\,, \\
\label{puretriplet_condition2}
\xi_\Delta \lambda_{H\Delta}-\xi_H \lambda_\Delta > & 0 \, .
\end{align}
\end{enumerate}
Notice that because of (\ref{pureHiggs_condition1}) and (\ref{puretriplet_condition2}) for pure Higgs and triplet inflation $\lambda_{H\Delta}>0$ is preferred.

\subsection{Slow-roll analysis for single field inflation}
\label{sec:inflPH}

\begin{figure}[t]
\begin{minipage}[t]{0.5\textwidth}
\centering
\includegraphics[width=0.9\columnwidth]{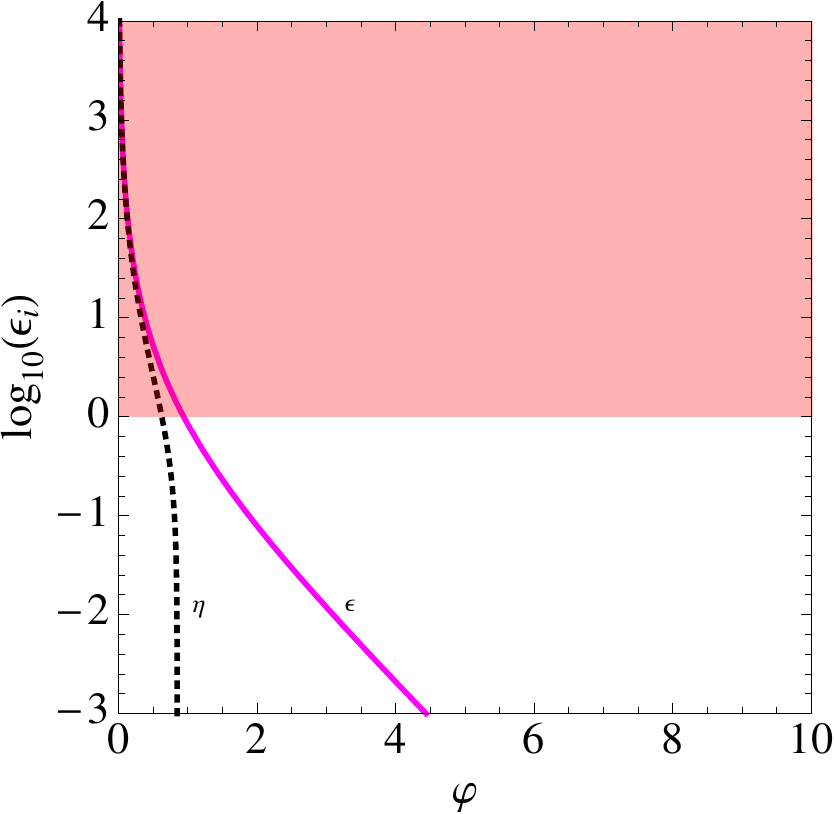}
\end{minipage}
\hspace*{-0.2cm}
\begin{minipage}[t]{0.5\textwidth}
\centering
\includegraphics[width=0.9\columnwidth]{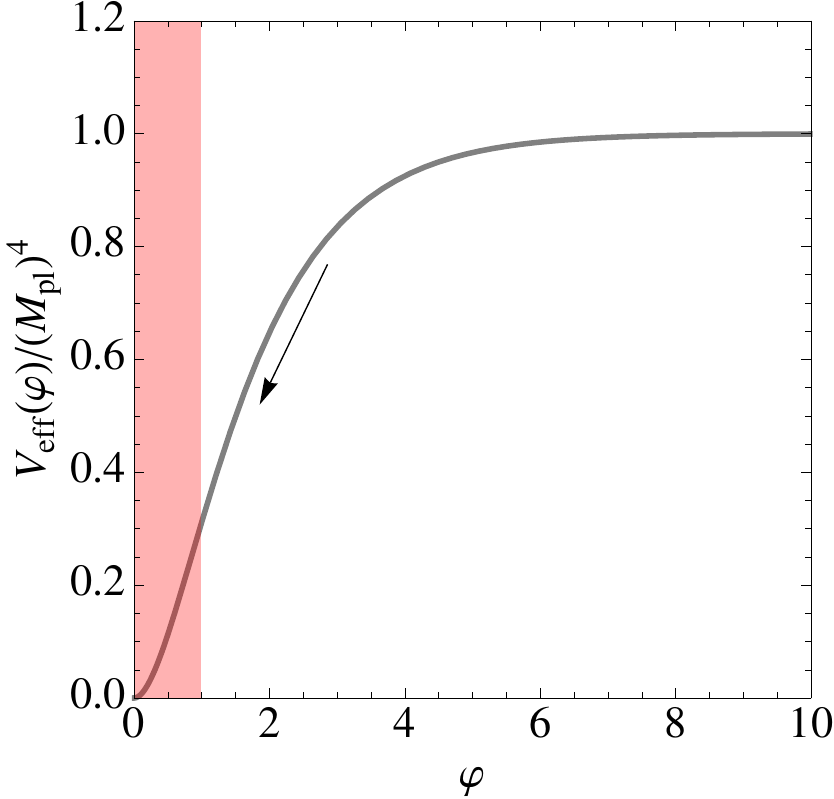}
\end{minipage}
\caption{{\it Left:} Two slow-roll parameters (\ref{epsilon}) and (\ref{eta}) are shown as a function of the inflaton field. The magenta solid line denotes $\epsilon$ while the black dotted line stands for $\eta$, as labeled. {\it Right:} The inflationary potential (\ref{eq:HiggsPot}) is depicted as a function of the inflaton, solid gray line. In both panels the light red region indicates where the slow roll condition $\epsilon<1$ break down.}
\label{fig:slowroll}
\end{figure}

Provided that the quartic potential alone is dominant over quadratic or triplet contributions to the potential, we may estimate the inflationary predictions using the so-called $\delta{N}$ formalism~\cite{Starobinsky:1986fxa,Sasaki:1995aw,Sasaki:1998ug,Gong:2002cx,Lyth:2005fi}. Essentially, the $\delta{N}$ formalism tells us that the perturbation in the number of $e$-folds, which is the same in both frames~\cite{Gong:2011qe}, is equivalent to the curvature perturbation on super-horizon scales. Then the slow-roll approximation, described by the parameters $\epsilon$ and $\eta$ is working well\footnote{Note that this approximation is equivalent at first order to the slow-roll predictions obtained with the Hubble flow parameters $\epsilon_n$, as described in~\cite{Martin:2006rs}.}.

Before going into the detail of slow-roll inflation let us make comments about the reheating. Inflation not only consists of the slow-roll period but also a reheating phase since it permits to link inflation with the subsequent radiation dominated era. This phase is connected to the potential part close to the minimum and takes place during a few $e$-folds. The reheating phase is poorly known and technically difficult to model properly. To take into account uncertainties on this post inflationary phase we use the reheating parameter $R_{\rm rad}$ described in~\cite{Martin:2010kz} as 
\begin{equation}\label{eq:reh}
\log R_{\rm rad} = \frac{\Delta N}{4} \left(-1 + 3 \overline{w}_{\rm reh} \right)\,,
\end{equation}
having supposed the simplest model of a scalar field coupled to radiation and that the effective fluid (inflaton plus radiation) with energy density $\rho$ and pressure $P$ is conserved and $\overline{w}_{\rm reh}$ stands for the mean equation of state parameter $w_{\rm reh}\equiv P/\rho$ during reheating. In addition $\Delta N$ is defined as the total number of $e$-folds during reheating
\begin{equation}\label{eq:Nreh}
\Delta N \equiv N_{\rm reh} - N_0\,,
\end{equation}
$N_{\rm reh}$ being the number of $e$-folds at which reheating is completed and the radiation dominated period begins while $N_0$ is the total number of $e$-folds during inflation. We assume instantaneous reheating, namely at the end of inflation the Universe enters straightaway in the radiation dominated era with equation of state $w=1/3$. This is equivalent to consider the reheating parameter equal to 1 or $\log R_{\rm rad} = 0$ in (\ref{eq:reh}). This can be understood physically because the pre-/reheating stage can not be distinguished from the radiation dominated era and therefore can not affect the inflationary predictions.

After the analysis of the previous sections the actual potential is, with $V_0 \equiv \lambda_{\rm eff}/\big(4\xi_{\rm eff}^2\big)$,
\begin{equation}\label{eq:HiggsPot}
V(\varphi) = V_0\,  \left(1-e^{-2\varphi/\sqrt{6}}\right)^2\,.
\end{equation}
Its behavior as a function of the field $\varphi$ is shown in the right panel of figure~\ref{fig:slowroll}. Note that inflation takes place for trans-Planckian values of the field. The shaded region denotes the breakdown of the slow-roll approximation, that we discuss straightaway. We can define the slow-roll parameters in terms of the potential as
\begin{align}
\label{epsilon}
\epsilon = & \frac{1}{2} \left( \frac{V'}{V} \right)^2 = \frac{4}{3}\frac{e^{-4\varphi/\sqrt{6}}}{\left( 1-e^{-2\varphi/\sqrt{6}} \right)^2} \, ,
\\
\label{eta}
\eta = & \frac{V''}{V} = -\frac{4}{3}e^{-2\varphi/\sqrt{6}} \frac{1-2e^{-2\varphi/\sqrt{6}}}{ \left( 1-e^{-2\varphi/\sqrt{6}} \right)^2} \, .
\end{align}
Both parameters, as functions of $\varphi$, are shown in the left panel of figure~\ref{fig:slowroll}, as labeled. Note that both are positive and monotonically increasing functions of the inflaton. The red region indicates the breakdown of slow-roll condition $\epsilon<1$. Then, the number of $e$-folds becomes, using the slow-roll equation $3H\dot\varphi+V'=0$,
\begin{equation}\label{efold}
N = \int_e^\star \frac{V}{V'} d\varphi = \frac{3}{4} \left[ e^{2\varphi_\star/\sqrt{6}} - e^{2\varphi_e/\sqrt{6}} - \frac{2}{\sqrt{6}}(\varphi_\star-\varphi_e) \right] \, .
\end{equation}
To determine the latter, we identify this moment as when $\epsilon(\varphi_e)=1$. Then we easily find
\begin{equation}\label{eq:bdsr}
\varphi_e = -\frac{\sqrt{6}}{2}\log\left(2\sqrt{3}-3\right) = 0.940 \, .
\end{equation}
The total number of $e$-folds after the Hubble length exit for instantaneous reheating is given by $\Delta N_{\star} = 55.6$, using (\ref{eq:reh}), enough to solve the flatness and horizon problems. We can find $\varphi_\star$ by plugging $\varphi_e$ into (\ref{efold}) as
\begin{equation}
\varphi_\star = 5.36 \, .
\end{equation}
From (\ref{efold}) we can immediately find
\begin{equation}\label{singledeltaN}
\frac{\partial{N}}{\partial\varphi_\star} = \frac{\sqrt{6}}{4} \left( e^{2\varphi_\star/\sqrt{6}} - 1 \right) = 48.3\, .
\end{equation}
The power spectrum  is normalized at the pivot scale of WMAP7, $k_0=0.002\,  {\rm Mpc^{-1}}$: $\calP_\calR(k_0)=(2.43 \pm 0.11) \times 10^{-9}$. The power spectrum of scalar perturbation at the pivot scale is given by:
\begin{equation}\label{eq:pptheo}
\calP_\calR(k_0) = \frac{V(\varphi_\star)}{12\pi^2} \left( \frac{\partial{N}}{\partial\varphi_\star} \right)^2 = \frac{V(\varphi_\star)}{24\pi^2\epsilon(\varphi_\star)} \, .
\end{equation}
Thus as quoted several times before, $\lambda_{\rm eff}$ is fixed once we fix $\xi_{\rm eff}$ or vice versa, as
\begin{equation}
\xi_{\rm eff} = \frac{\sqrt{\lambda_{\rm eff}}}{\sqrt{96\pi^2\epsilon(\varphi_\star)\calP_\calR(k_0)}}
= 48646.2\sqrt{\lambda_{\rm eff}} \sim 5\times10^4 \sqrt{\lambda_{\rm eff}} \, .
\end{equation}
Also, the spectral index $n_\calR$ is given by
\begin{equation}
n_\calR = 1-2\epsilon+2\eta-\frac{2}{(\partial{N}/\partial\varphi_\star)^2} = 0.965 \, ,
\end{equation}
which lies well within the 2$\sigma$ range of WMAP7, $n_\calR = 0.963 \pm 0.0014$~\cite{Komatsu:2010fb}.

\section{Asymmetric DM and Leptogenesis}
\label{sec:ADML}

From (\ref{eq:Lag-DM}) we see that there are three different channels available for the decay of scalar triplet $\Delta$: $\Delta \to LL$, $\Delta \to HH$ and $\Delta \to \psi \psi$. Since the couplings are in general complex, the quasi-equilibrium decay of $\Delta$ via these channels produce asymmetries in lepton and DM sectors.  
\begin{figure}[t]
\centering
\includegraphics[width=.8\columnwidth]{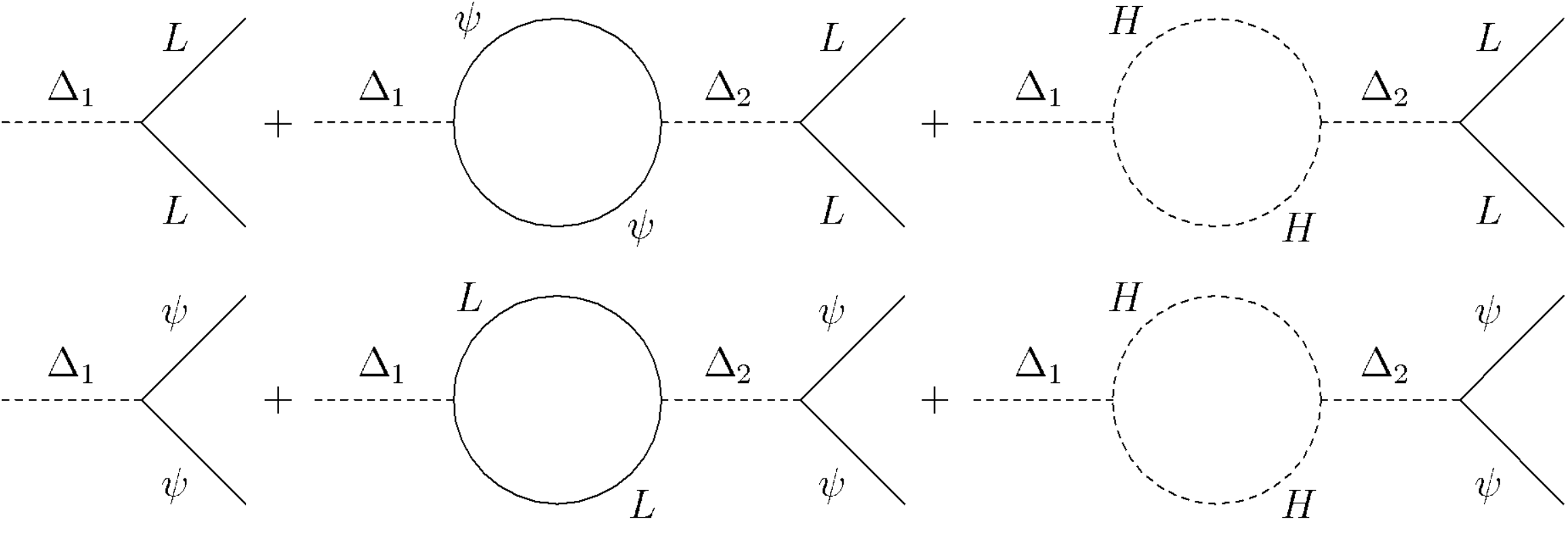}
\caption{The CP asymmetry for leptogenesis and asymmetric DM generated by the interference of tree and one-loop self energy correction diagrams are shown in the first and second raw respectively.}
\label{cpasy_vis}
\end{figure} 
The CP asymmetry in either sector arises via the interference of tree level with the one-loop self energy diagram as shown in figure~\ref{cpasy_vis}. From these diagrams we see that to generate net CP asymmetry at least two scalar triplets $\Delta_1$ and $\Delta_2$ are required. As a result the interaction of $\Delta_1$ and $\Delta_2$ is described by a complex mass matrix instead of a single mass term as mentioned in (\ref{eq:ScalarPotential}). The diagonalisation of the flavour basis spanned by ($\Delta_1, \Delta_2$) gives rise to two mass eigenstates $\zeta^+_{1,2} = A^+_{1,2} \Delta_1 + B^+_{1,2} \Delta_2$ with masses $M_1$ and $M_2$. The complex conjugate of $\zeta^+_{1,2}$ are given by $\zeta^-_{1,2} = A^-_{1,2} \Delta_1 + B^-_{1,2} \Delta_2$. Unlike the flavour eigenstates $\Delta_1$ and $\Delta_2$, the mass eigenstates $\zeta^+_{1,2}$ and $\zeta^-_{1,2}$ are not CP eigenstates and hence their decay can give rise to CP asymmetry. Assuming a mass hierarchy in the mass eigenstates of the triplets, the final asymmetry arises by the decay of lightest triplet $\zeta_1^+$ and $\zeta_1^-$. The CP asymmetries are defined as
\begin{align}
\epsilon_L =&\, \,  2 \left[ {\rm Br}(\zeta_1^{-}\to \ell \ell) - {\rm Br} (\zeta_1^{+} \to
\ell^c \ell^c) \right] \,,  
\\
\epsilon_{\psi} =&\,\,   2 \left[ {\rm Br} \left( \zeta_1^{-}\to \psi_{\rm DM} \psi_{\rm DM} \right) 
- {\rm Br} \left( \zeta_1^{+}
\to \psi_{\rm DM}^c \psi_{\rm DM}^c \right) \right] 
\equiv \epsilon_{\rm DM} \,,
\end{align}
where the front factor 2 takes into account of two similar particles are produced per decay. From figure~\ref{cpasy_vis}, the asymmetries are estimated to be:
\begin{align}
\epsilon_{L} = & \, \, \frac{1}{8\pi^2} \frac{M_1 M_2}{M_2^2-M_1^2} \left[\frac{M_1}{\Gamma_{1}} \right]
  {\rm Im}\left[  \left( f_{1\psi} f_{2\psi}^* + f_{1H} f_{2H}^*\right) \sum_{\alpha \beta} 
(f_{1L})_{\alpha\beta} (f^*_{2L})_{\alpha\beta} \right]\,,
\label{cp_vis}
\\
\epsilon_{\rm DM} = &\, \,  \frac{1}{8\pi^2} \frac{M_1 M_2}{M_2^2-M_1^2} \left[\frac{M_1}{\Gamma_{1}} \right] 
{\rm Im} \left[ f_{1\psi} f_{2\psi}^* 
\left( f_{1H} f_{2H}^* + \sum_{\alpha \beta} (f_{1L})_{\alpha\beta} (f^*_{2L})_{\alpha\beta}
\right) \right] \,,
\label{cp_dark}
\end{align}
where 
\begin{equation}
\Gamma_1=\frac{M_1}{8\pi} \left( |f_{1H}|^2 + |f_{1\psi}|^2 + |f_{1L}|^2 \right)\,,
\end{equation}
is the total decay rate of the lightest triplet. In the numerical calculations we will use this total decay rate as: $\Gamma_1(m_\nu, B_L, B_H, M_1)$, where $B_L$ and $B_H$ are the branching fractions in the decay channels $\zeta_1 \to L L$ and $\zeta_1 \to H H$ respectively. In the following we set $m_\nu=0.05$ eV and therefore the total decay rate depends only on three variables, namely $B_L$, $B_H$ and $M_1$.

When $\Gamma_1$ fails to compete with the Hubble expansion scale of the Universe, $\zeta_1$ decays away and produces asymmetries in either sectors. As a result the yield factors are given by:
\begin{align}
Y_L  \equiv &\, \,  \frac{n_L}{s} = \epsilon_L X_\zeta \eta_L\,, 
\\
Y_{\rm DM} \equiv &\, \,  \frac{n_\psi}{s}  =  \epsilon_{\rm DM} X_\zeta \eta_{\rm DM}\,,
\label{asymmetry_L_DM}
\end{align}
where $X_\zeta = n_{\zeta_1^-}/s \equiv n_{\zeta_1^+}/s$, $s=2\pi^2 g_* T^3/45$ is the entropy density and $\eta_{L}$, $\eta_{\rm DM}$ are the efficiency factors, which take into account the depletion of asymmetries due to the number violating processes involving $\psi$, $L$ and $H$. At a temperature above EW phase transition a part of the lepton asymmetry gets converted to the baryon asymmetry via the $SU(2)_L$ sphaleron processes. As a result the baryon asymmetry is:
\begin{equation}
Y_B = -\frac{8 n + 4 m}{14 n + 9 m} Y_L = -0.55 Y_L\,,
\label{B-asy}
\end{equation}
where $n$ is the number of generations and $m$ is the number of scalar doublets. From (\ref{asymmetry_L_DM}) and (\ref{B-asy}) the DM to baryon ratio is given by:
\begin{equation}\label{eq:IMP}
\frac{\Omega_{\rm DM}}{\Omega_B} = \frac{1}{0.55 }\frac{m_{\rm DM}}{m_p} \frac{\epsilon_{\rm DM}}{\epsilon_L}
\frac{\eta_{\rm DM}}{\eta_L}\,,
\end{equation}
where $m_p\sim 1\GeV $ is the proton mass. From this equation it is clear that the criteria $\Omega_{\rm DM} \sim 5\  \Omega_B$ can be satisfied by adjusting the ratio $\epsilon_{\rm DM}/\epsilon_L$ and $\eta_{\rm DM}/\eta_L$, where the efficiency factor:
\begin{equation}
\eta_i = \frac{Y_i}{\epsilon_i \ X_\zeta\Big|_{T \gg M_1}} \quad \text{with} \quad i={\rm DM, }~L
\end{equation}
can be obtained by solving the relevant Boltzmann equations~\cite{Arina:2011cu,Hambye:2005tk,Chun:2006sp} given in~\ref{appA}. The ratio of CP asymmetries is
\begin{equation}
\frac{\epsilon_{\rm DM}}{\epsilon_L} = \frac{{\rm Im} \left[ f_{1\psi} f_{2\psi}^*
\left( f_{1H} f_{2H}^* + \sum_{\alpha \beta} (f_{1L})_{\alpha\beta} (f^*_{2L})_{\alpha\beta}
\right) \right] }{{\rm Im}\left[  \left( f_{1\psi} f_{2\psi}^* + f_{1H} f_{2H}^*\right) \sum_{\alpha \beta}
(f_{1L})_{\alpha\beta} (f^*_{2L})_{\alpha\beta} \right]} \, .
\end{equation} 
From the above equation we observe that if $f_\psi > f_H \gg f_L$, then we get 
\begin{equation}
\frac{\epsilon_{\rm DM}}{\epsilon_L} \sim  \frac{{\cal O} (f_H^2)}{{\cal O} (f_L^2)}\,.
\label{CP-ratio}
\end{equation}
Taking $10^{-5} < f_H < 0.1$ and $f_L \sim 10^{-5}$ we get the ratio of CP asymmetries in a broad range: $10^2 - 10^8$.

\section{Renormalisation Group Equations in Scalar Triplet Model}
\label{sec:RGEs}

The RG equations of the scalar, gauge and Yukawa couplings in SM have been extensively discussed in the literature, see for example~\cite{DeSimone:2008ei,Espinosa:2007qp} for discussions relative to the cosmological framework. However, in the presence of scalar triplet the RG evolution of these couplings change because of the additional lepton number violating interactions of the scalar triplet with SM Higgs and leptons, as it has been described first in~\cite{Chao:2006ye} and then improved by~\cite{Schmidt:2007nq,Gogoladze:2008gf}, which will be our main references. Moreover, in our case, the triplet couples to the inert fermion doublet dark matter $\psi$. In the following we list the modification to the standard running for $\lambda_H$ as well as the RG equations for the different couplings pertaining to $\Delta$ such as $\lambda_\Delta$, $\lambda_{H\Delta}$, $\mu_H$, $f_\psi$ and the non-minimal couplings to gravity $\xi_\Delta$ and $\xi_H$.

Having defined $\beta_X= dX/d\ln\mu$, where $\mu$ is the renormalization scale, the RG equations of the quartic couplings in the scalar potential including the triplet are given by
\begin{align}
\label{eq:RGquartic}
16\pi^2 \beta_{\lambda_H} = & 12\lambda_H^2 + 6 \lambda^2_{H\Delta}- \left( \frac{9}{5} g_1^2 + 9 g_2^2 \right)\lambda_H 
 + \frac{9}{4} \left(\frac{3}{25} g_1^4 + \frac{2}{5} g_1^2 g_2^2 + g_2^4 \right) + \left( 12 \lambda_H Y_t^2 - 12 Y_t^4\right) \,,
\\
16\pi^2 \beta_{\lambda_\Delta}  =&  -\left(\frac{36}{5}g_1^2 + 24 g_2^2 \right)\lambda_\Delta
+ \frac{108}{25} g_1^4 + 18 g_2^4 + \frac{72}{5} g_1^2 g_2^2  + 14 \lambda_\Delta^2 + 4 \lambda_{\Delta H}^2 
\nonumber\\
& + 4 \lambda_{\Delta} {\rm Tr} \left(f_L^\dagger f_L + f_\psi^\dagger f_\psi \right)
- 8 {\rm Tr}\left(f_L^\dagger f_L f_L^\dagger f_L + f_\psi^\dagger f_\psi f_\psi^\dagger f_\psi\right)\,,
\\ 
16\pi^2  \beta_{\lambda_{\Delta H}} =& -\left(\frac{9}{2}g_1^2 + \frac{33}{2} g_2^2 \right)
\lambda_{\Delta H} +  \frac{27}{25} g_1^4 + 6 g_2^4   + \left(8 \lambda_\Delta + 6\lambda_H + 4 \lambda_{\Delta H} + 6 Y_t^2 \right)\lambda_{\Delta H} 
\nonumber\\
& + 2 {\rm Tr}\left(f_L^\dagger f_L + f_\psi^\dagger f_\psi \right) \lambda_{\Delta H}  - 4 {\rm Tr}\left(f_L^\dagger f_L f_L^\dagger f_L + f_\psi^\dagger f_\psi f_\psi^\dagger f_\psi \right) \,,
\end{align}
where we have assumed that the dominant contribution to RG evolution comes from the top quark Yukawa coupling $Y_t$ in the SM. Note that the beta function for $\lambda_H$ gets a positive contribution from $\lambda_{H\Delta}$, whose importance will be discussed later on. $g_1$, $g_2$ and $g_3$ are the couplings corresponding to the gauge groups $U(1)_Y$, $SU(2)_L$ and $SU(3)_C$ of the SM. In presence of the scalar triplet the evolution of gauge couplings are given by:
\begin{align}
16 \pi^2 \beta_{g_1} =&  \frac{47}{10}\, g_1^3\,, \\
16 \pi^2 \beta_{g_2} =&  -\frac{5}{2}\, g_2^3\,, \\
16 \pi^2 \beta_{g_3} =&  -7\, g_3^3 \,.
\end{align}
Since the triplet is a singlet under $SU(3)_C$, the running of $g_3$ is not affected. By the same argument the running of the Yukawa coupling for top quark is not affected either:
\begin{equation}
16\pi^2 \beta_{Y_t} = \frac{9}{2} Y_t^3 -\left(\frac{17}{20} g_1^2 + \frac{9}{4}g_2^2 + 8 g_3^2 \right) Y_t\,.
\end{equation}
The RG evolution of the Majorana Yukawa couplings $f_L$ and $f_\psi$ are given by:
\begin{align}
16\pi^2 \beta_{f_L} =&  3 \left( f_L^\dagger f_L + f_\psi^\dagger f_\psi \right) f_L -
\frac{3}{2} \left(\frac{3}{5} g_1^2 + 3 g_2^2 \right) f_L + \left[ {\rm Tr}\left(f_L^\dagger f_L 
+ f_\psi^\dagger f_\psi \right) \right] f_L\,, 
\\
16\pi^2 \beta_{f_\psi} =&  3 \left( f_L^\dagger f_L + f_\psi^\dagger f_\psi \right) f_\psi 
-\frac{3}{2} \left(\frac{3}{5} g_1^2 + 3 g_2^2 \right)f_\psi + \left[ {\rm Tr}\left(f_L^\dagger f_L 
+ f_\psi^\dagger f_\psi \right) \right] f_\psi \,.
\end{align}
The RG equation of the coefficient of the trilinear $\Delta^\dagger HH$ coupling is given by
\begin{equation}
16\pi^2 \beta_{\mu_H}  =  \left( \lambda_H - 4 \lambda_{\Delta H} -\frac{27}{10} g_1^2 -\frac{21}{2} g_2^2 
+ 6 Y_t^2 \right) \mu_H  +\left[  {\rm Tr} \left(f_L^\dagger f_L + f_\psi^\dagger f_\psi \right) 
\right] \mu_H\,.
\end{equation}
The anomalous dimensions of $\Delta$ and $H$ are 
\begin{align}\label{eq:adim}
16\pi^2 \gamma_{M_\Delta} =& \frac{9}{5} g_1^2 + 6 g_2^2 -4 \lambda_\Delta - {\rm Tr} \left(f_L^\dagger f_L 
+ f_\psi^\dagger f_\psi \right) 
 - \left(2 \lambda_{\Delta H} \frac{M_H^2}{M_\Delta^2} + \frac{1}{2}\frac{|\mu_H|^2}{M_\Delta^2}  \right)\,, 
\\
16\pi^2 \gamma_{M_H} =& \frac{9}{20} g_1^2 + \frac{9}{4} g_2^2 - \frac{3}{2} \lambda_H - 
3 Y_t^2  - 3 \left( \lambda_{\Delta H} \frac{M_\Delta^2}{M_H^2} + \frac{|\mu_H|^2}{M_H^2}\right)\,,
\end{align}
where the symbol $\gamma_X$ is defined by $\gamma_X\equiv -X^{-1}dX/d\ln\mu$. The anomalous dimensions have a key role in determining the RG equations for the non-minimal couplings $\xi_H$ and $\xi_\Delta$. As described in~\cite{Buchbinder:1992rb}, for a general theory of scalars $\phi_i$ coupled non-minimally to gravity via $\xi_{ij}$, the one-loop bare and renormalized non-minimal couplings are defined as
\begin{equation}
\xi_{0ij} = \left( \xi_{kl} - \frac{1}{6} \delta_{kl} \right) Z_{2ij}^{kl} + \frac{1}{6} \delta_{ij}\,,
\end{equation}
where $Z_{2ij}^{kl}$ denotes the mass renormalization term $m^2_{0ij} = Z_{2ij}^{kl} m^2_{kl}$. The RG equations for the $\xi_{ij}$ are linked to the mass anomalous dimensions by
\begin{equation}\label{eq:defxirge}
\beta_{\xi_{ij}} = \left( \xi_{mn}-\frac{1}{6} \delta_{mn}\right) \gamma^{kl}_{ij}\,.
\end{equation}
Consequently, plugging (\ref{eq:adim}) into the definition~(\ref{eq:defxirge}), the beta function for the non-minimal couplings are
\begin{align}
16 \pi^2 \beta_{\xi_H} = & \left(\xi_H +\frac{1}{6}\right) \left(- \frac{9}{20} g_1^2 - \frac{9}{4} g^2_2 
+ \frac{3}{2}\lambda_H + 3 Y_t^2\right) + 3 \left(\xi_\Delta +\frac{1}{6}\right) \left(\lambda_{H\Delta} 
+  \frac{\mu_H^2}{M_\Delta^2}\right)\,,
\\
16 \pi^2 \beta_{\xi_\Delta} = &  2 \left(\xi_H +\frac{1}{6}\right) \lambda_{H\Delta} + \left(\xi_\Delta 
+\frac{1}{6}\right) \left[4 \lambda_\Delta + \frac{1}{2}\frac{\mu_H^2}{M_\Delta^2} 
+ {\rm Tr} \left(f_L^\dagger f_L + f_\psi^\dagger f_\psi \right) -\frac{9}{5}g_1^2 - 6g_2^2 \right]\,.
\end{align}

For renormalization scales below $M_\Delta$, the triplet is decoupled and should be integrated out. Therefore, the set of equations reduces to the SM ones with one important modification. Indeed in the decoupling limit ($\mu < M_\Delta$) we see that the trilinear $\Delta^\dagger H H$ term provides an effective term 
\begin{equation}
-\frac{1}{2} \left(\frac{\mu_H^\dagger \mu_H}{M_\Delta^2} \right) (H^\dagger H)^2\,.
\end{equation} 
As a result the effective quartic coupling of the SM Higgs is modified as
\begin{equation}\label{eq:lambdaH}
\Lambda= \lambda_H -\frac{1}{2} \left(\frac{\mu_H^\dagger \mu_H}{M_\Delta^2} \right)\,,
\end{equation}
with a beta function equivalent to the SM one following the prescription\footnote{We will use both $\Lambda$ and $\lambda_H$ to indicate the Higgs quartic coupling. It should be intended that in the decoupling limits the two are strictly equivalent, while above the triplet mass scale both notations refer to $\lambda_H$.} $\lambda_H \to \Lambda$:
\begin{equation}
16\pi^2 \beta_{\Lambda}  =  12\Lambda^2 - \left( \frac{9}{5} g_1^2 + 9 g_2^2 \right)\Lambda
 + \frac{9}{4} \left(\frac{3}{25} g_1^4 + \frac{2}{5} g_1^2 g_2^2 + g_2^4 \right) 
+ \left( 12 \Lambda Y_t^2 - 12 Y_t^4\right) \,.
\end{equation}
In the decoupling limit, the beta functions of the gauge couplings are also given by 
\begin{align}
16 \pi^2 \beta_{g_1} =&\frac{41}{10}\, g_1^3\,, \\
16 \pi^2 \beta_{g_2} =& - \frac{19}{6} \,g_2^3\,, \\
16 \pi^2 \beta_{g_3} =& -7\, g_3^3\,. 
\end{align}
The RG equations above and below the mass scale of the triplet are matched at $M_\Delta$, with particular care of the condition in (\ref{eq:lambdaH}) for the quartic coupling of the Higgs. For the initial conditions at EW scale we use the renormalization scale $\mu = m_t$ as suggested in~\cite{Hambye:1996wb}, with $m_t=172.9$ GeV~\cite{0954-3899-37-7A-075021} and $v=246.22$ GeV. The gauge coupling constants are fixed at the following values: $\alpha_1(m_t)=0.01027$, $\alpha_2(m_t)=0.03344$ and $\alpha_3(m_t)=0.1071$. We use the pole matching scheme for $\lambda_H(m_t)$ and $Y_t(m_t)$ as detailed in~\cite{Gogoladze:2008gf} and references therein, to relate the physical pole masses to the couplings in the $\overline{\rm MS}$ renormalization scheme. The free parameters $\lambda_\Delta$, $\lambda_{H\Delta}$, $f_\Delta$, $f_\psi$ and $\mu_H$ are fixed as well at $\mu = m_t$, the only difference being that their running will start only at $\mu = M_\Delta$. Below $M_\Delta$ only $\Lambda$ will have an effect on the $\xi_i$, fixed at $m_t$ as well. The running of the coupling is stopped at the unitarity scale $\mu_U= \min (M_{\rm pl}/\xi_H,M_{\rm pl}/\xi_\Delta)$.
\begin{figure}[t]
\begin{minipage}[t]{0.5\textwidth}
\centering
\includegraphics[width=1.\columnwidth,trim=0mm 15mm 15mm 30mm, clip]{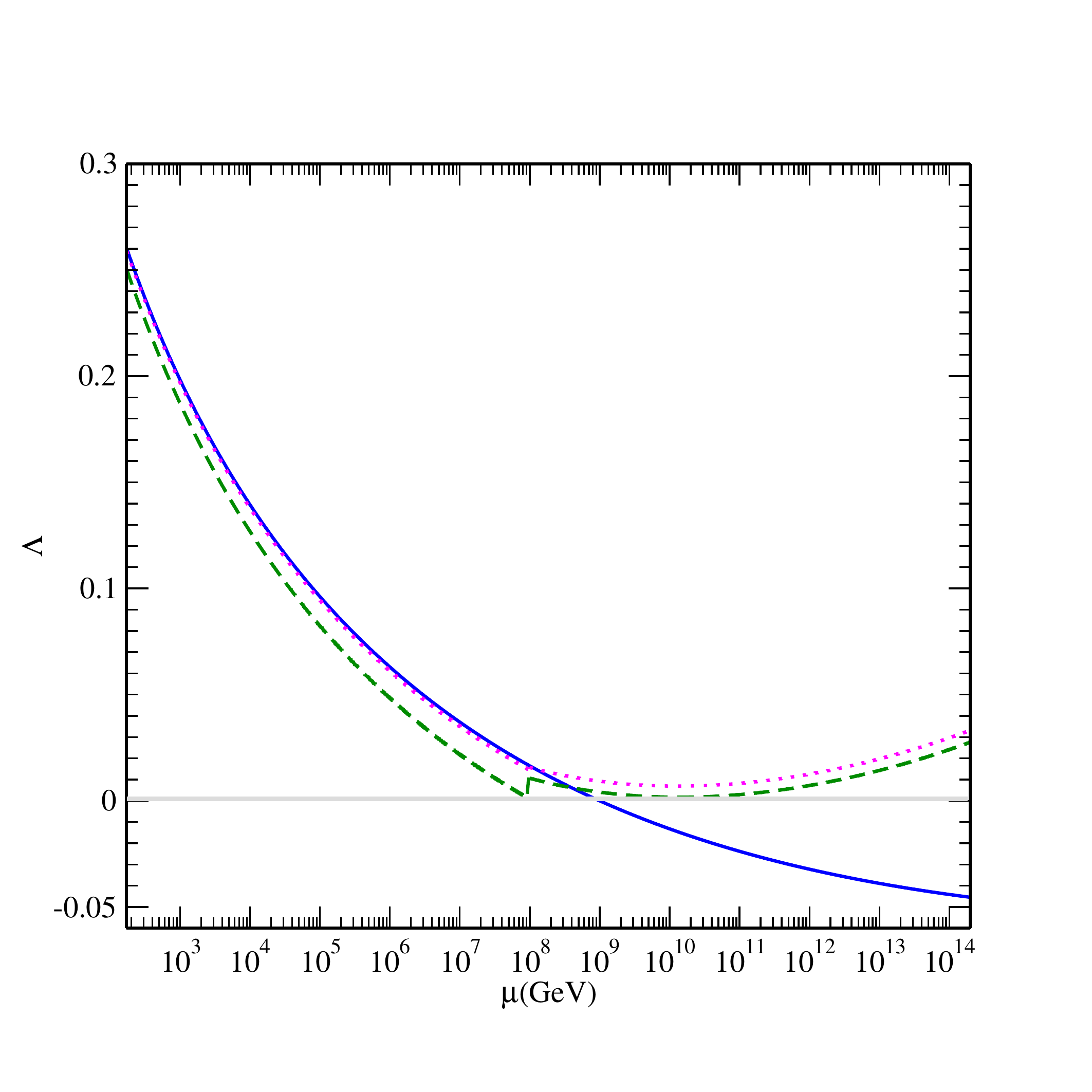}
\end{minipage}
\hspace*{-0.2cm}
\begin{minipage}[t]{0.5\textwidth}
\centering
\includegraphics[width=1\columnwidth,trim=0mm 15mm 15mm 30mm, clip]{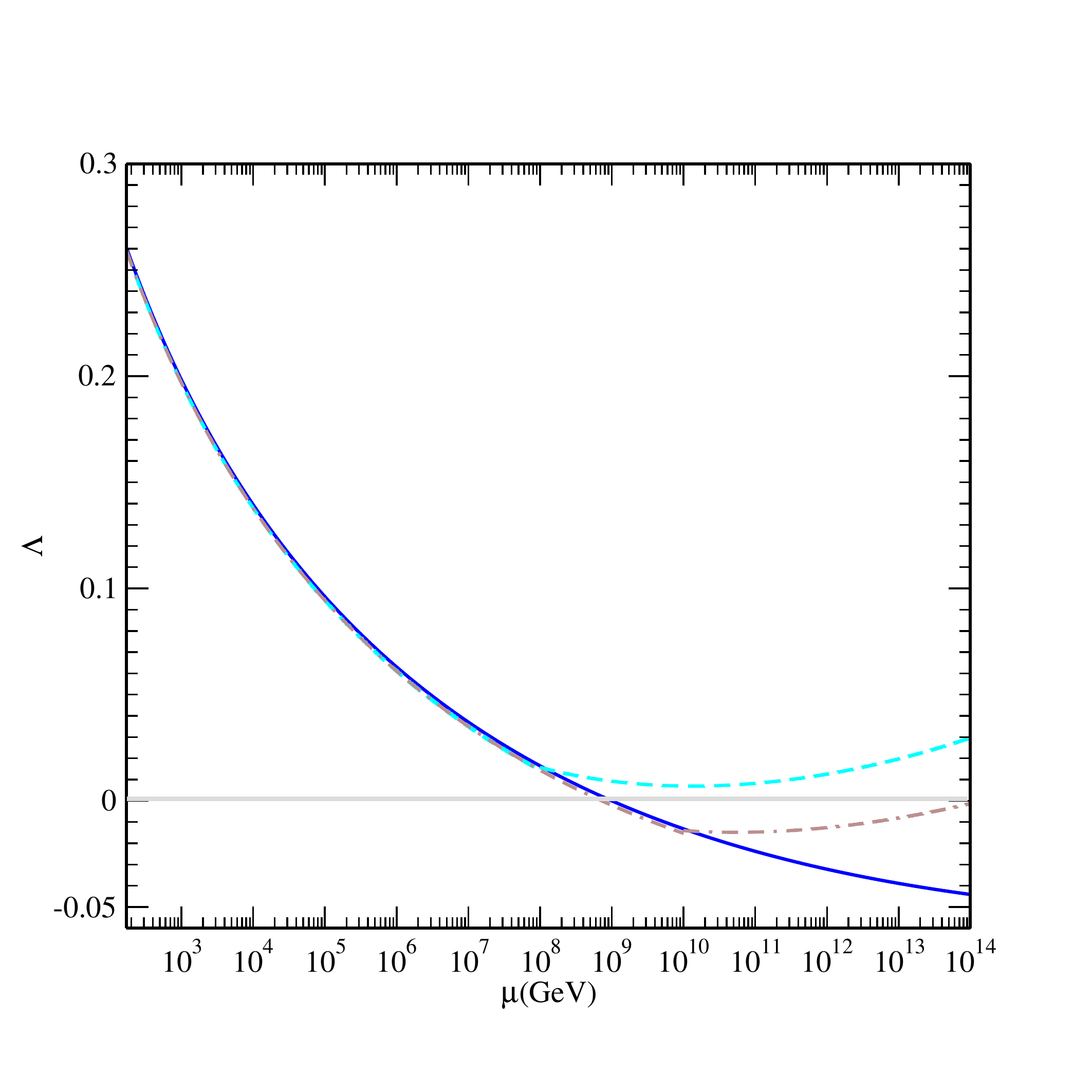}
\end{minipage}
\caption{Evolution of the Higgs quartic coupling $\Lambda$ as a function of the renormalization scale $\mu$ until the unitarity scale. In both panels, the blue solid line denotes the SM evolution with $\lambda_H \equiv \Lambda$ and the horizontal gray line denotes the metastability bound. The Higgs mass is fixed at $M_H=125$ GeV and hence $\lambda_H=0.26$ at the matching scale $\mu=m_t$. The common triplet parameters are $\lambda_\Delta=0.4$ and $\lambda_{H\Delta}=0.33$, $f_L=f_\psi=0.01$ and $\xi_\Delta=\xi_H=10^4$. {\it Left, effect of $\mu_H$}: the green dashed curve stands for $\mu_H=0.1 M_\Delta$, while the magenta dotted line is for $\mu_H=0.05 M_\Delta$. In both cases the triplet mass is $M_\Delta=10^8$ GeV. {\it Right, effect of $M_\Delta$}: the cyan dashed line denotes the running for $M_\Delta = 5\times 10^8$ GeV, while the brown dot-dashed line is for $M_\Delta=10^{10}$ GeV. Both curves have $\mu_H$ fixed at $0.01 M_\Delta$.}
\label{fig:RGeH}
\end{figure}

The Higgs mass is taken to be 125 GeV, the value at which CMS~\cite{CMS} has reported an excess for a Higgs like particle with a significance of about $5 \sigma$ (and Tevatron~\cite{TEVNPH:2012ab} with a smaller significance), similarly to the $5 \sigma$ significance of the Higgs boson like found by ATLAS~\cite{ATLAS} at around 126 GeV. The Higgs quartic coupling will then have a definite value ($\lambda_H = 0.26$ at $m_t$) through the matching pole condition. The running of this coupling in the SM is then fixed and goes negative at around $10^9$ GeV. A possible way of avoiding such a metastable vacuum of the SM is to introduce the scalar triplet at that scale. As a result the positive correction proportional to $\lambda_{H\Delta}$ in (\ref{eq:RGquartic}) prevent the Higgs quartic coupling to run towards negative values. The role played by the triplet is illustrated in figure~\ref{fig:RGeH}. The blue solid line stands for the running of $\lambda_H$ in the SM: it is clear that at $10^9$ GeV this coupling goes below zero, denoted by the gray horizontal line. Additionally to this curve, in the left panel we show the running of $\Lambda$ for different values of $\mu_H$ taking $M_\Delta=10^8$ GeV. Since the contribution of $\mu_H$ in (\ref{eq:lambdaH}) is always negative, the larger its value the sooner $\Lambda$ becomes negative: the magenta dotted curve is for $f_H=0.05$, while the green dashed line stands for $f_H=0.1$, which is the upper bound on this parameter in order to prevent instability below $10^8$ GeV. The small step is due to the matching condition at $M_\Delta$ of $\Lambda$ and $\lambda_H$. In the right panel, together with the SM running of $\lambda_H$, we illustrate the role of the triplet mass: the cyan dashed line shows that taking the triplet mass at $5 \times  10^8$ GeV and $\lambda_{H\Delta}$ at least larger than 0.33 will permit the quartic coupling of the Higgs to stay positive until unitarity scale, at which it has the value of $\mathcal{O}(0.1)$. On the contrary, if the triplet mass scale is too high, as shown by the brown dot-dashed line with $M_\Delta=10^{10}$ GeV, the theory can not be rescued from vacuum instability. In the above discussions, the values of other parameters can be read from caption.

Assuming the Higgs mass is fixed at $125$ GeV, the following conclusions can be drawn:
\begin{enumerate}

\item {\it Effect of $\mu_H$}: The contribution of $\mu_H$ to $\Lambda$ is always negative, therefore it lowers the instability threshold. In order not to loose the interest of the seesaw mechanism to produce neutrino masses in the sub-eV range, we keep the mass of the triplet at around $10^{8}$ GeV. In addition the ratio $f_H \equiv \mu_H/M_\Delta$ gives the neutrino mass via type-II seesaw, which can not be negligibly small, hence $f_H$ should be at least $\mathcal{O}(10^{-2})$. We note that those values of $f_H$ have negligible effect on the running of $\lambda_H$ and hence on the Higgs mass (a large $\mu_H$ could shift the Higgs mass via the tree level relation $M_H=\sqrt{\Lambda} v$) however may not be too small for neutrino masses.

\item {\it Effect of $M_\Delta$}: In order to avoid instability of the potential, the mass of the triplet should be lower than the instability point. Therefore, masses of the triplet larger than $ 7 \times 10^{8}$ GeV are disfavored. In the following we fix $M_\Delta = 10^8$ GeV, which is still in the ballpark to get the correct sub-eV neutrino masses, with a corresponding upper bound on $f_H$ of $\mathcal{O}(0.1)$.

\end{enumerate}
\begin{figure}[t!]
\begin{minipage}[t]{0.5\textwidth}
\centering
\includegraphics[width=1.\columnwidth,trim=0mm 15mm 15mm 30mm, clip]{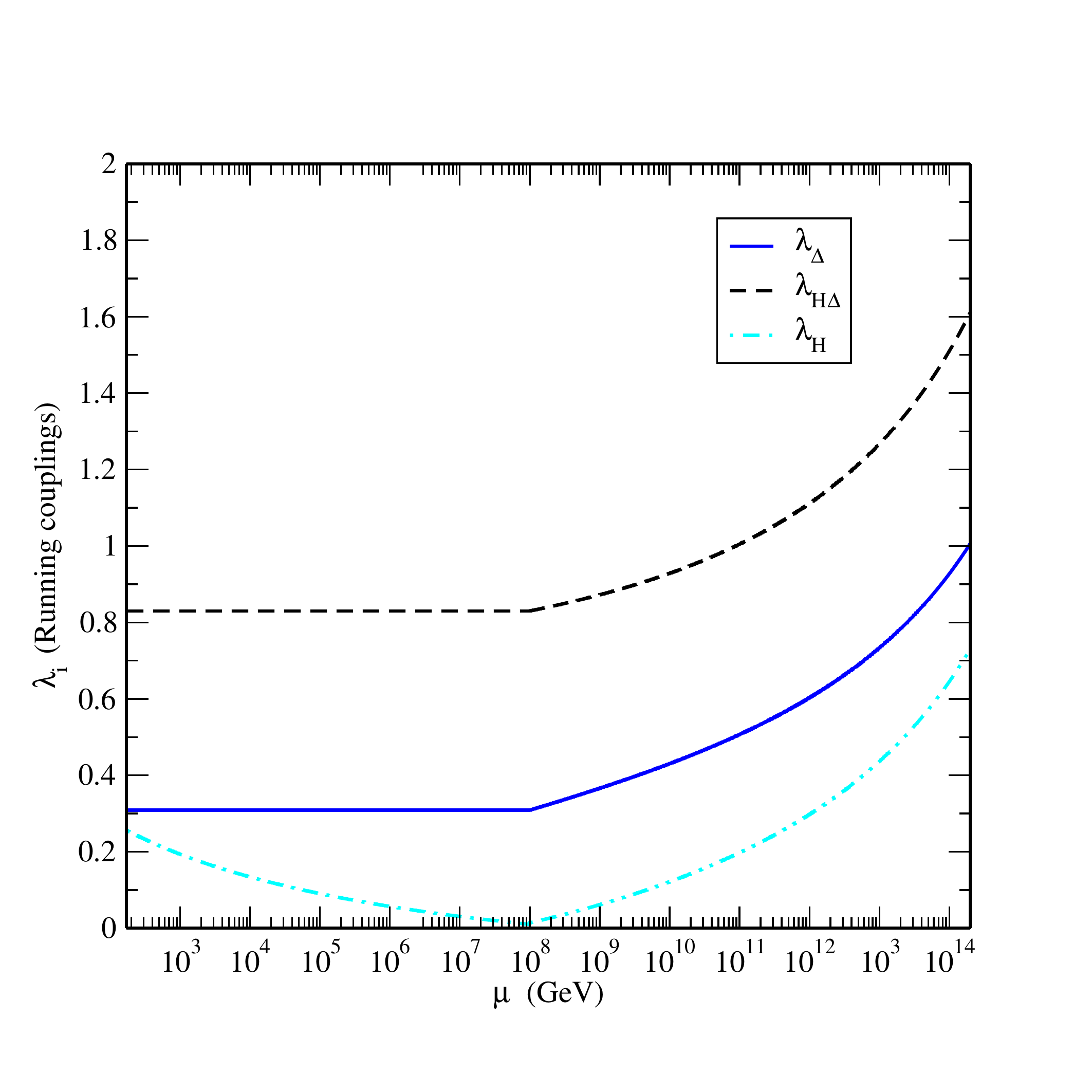}
\end{minipage}
\hspace*{-0.2cm}
\begin{minipage}[t]{0.5\textwidth}
\centering
\includegraphics[width=1.\columnwidth,trim=0mm 15mm 15mm 30mm, clip]{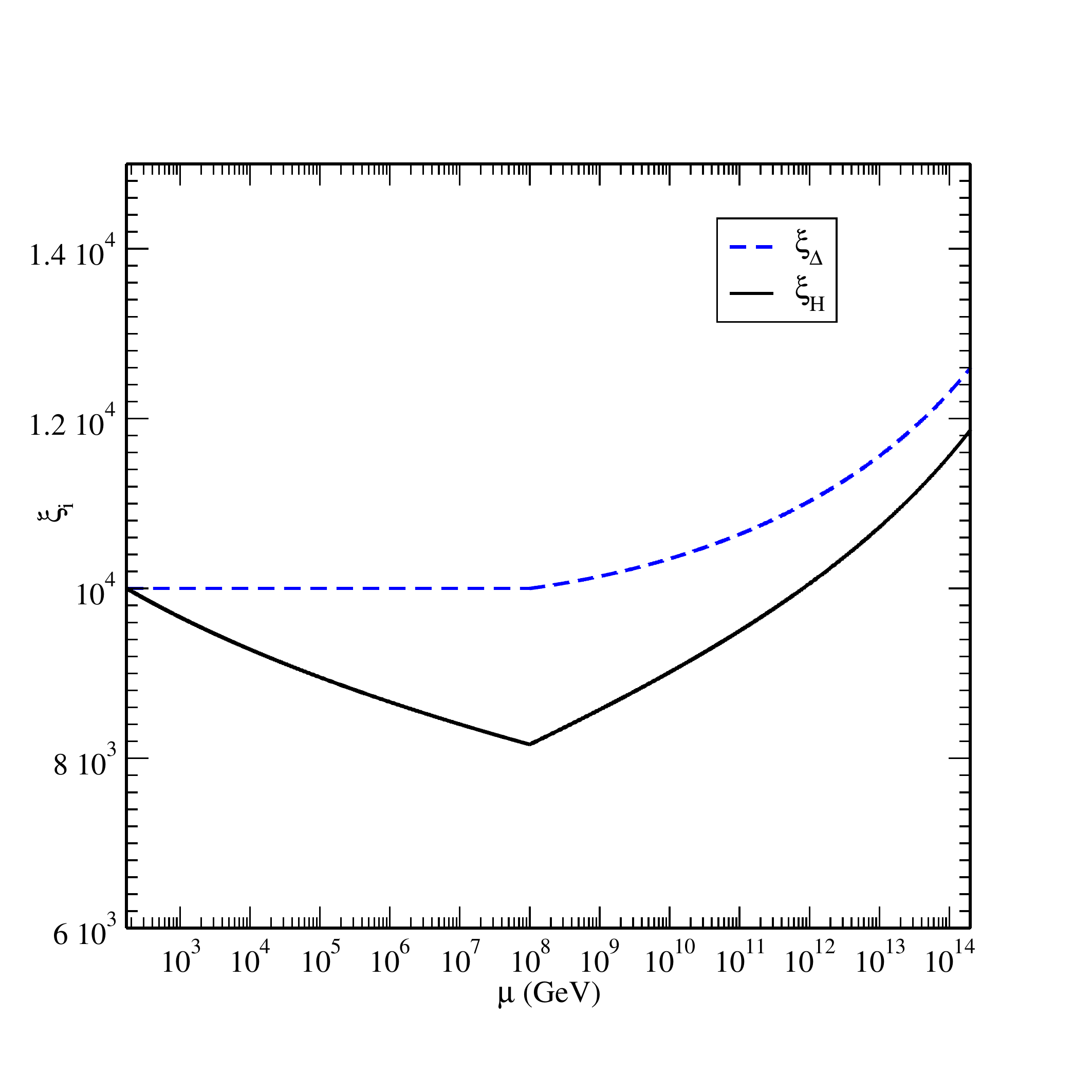}
\end{minipage}
\caption{An example of RG evolution as a function $\mu$ from $m_t$ to the unitary scale $\mu_U$. The fixed parameters are: $M_H=125$ GeV, $\lambda_H=0.26$, $f_L=f_\psi=0.5$, $\mu_H/M_\Delta=0.09$. {\it Left:} Running of the coupling constants in the scalar potential as labeled in the plot. {\it Right:} RG evolution of the non-minimal couplings to gravity as labeled.}
\label{fig:exRGE}
\end{figure}

In the left panel of figure~\ref{fig:exRGE}, we show an example of RG running of various quartic couplings in the scalar potential, while on the right panel we show the running of non-minimal couplings to gravity, as labeled in the plots. Below the mass scale of triplet $M_\Delta$ the quartic couplings $\lambda_{H\Delta}$ and $\lambda_\Delta$ remain constant and start to running above $M_\Delta$ due to the globally positive correction from $g_i$ and $f_i$. The non-minimal coupling of the Ricci scalar with the Higgs receives a negative contribution from the SM parameters below the triplet mass scale, while above $M_{\Delta}$ the corrections due to the triplet parameters set on and increase the value of $\xi_H$, which ends up to be larger than its EW value. On the other hand, $\xi_\Delta$ does not receive any contribution from the SM parameters and therefore remains constant up to $M_\Delta$ and then increases.

We note that for numerical purpose, we use the RG equations with two-loop corrections for SM variables while one-loop level corrections for the running of triplet scalar couplings, as described in~\cite{Gogoladze:2008gf}.

\section{Results and Discussions}
\label{sec:res}

In this section we discuss the model parameter space which satisfy the inflationary, DM and leptogenesis constraints. The following points are in order:  
\begin{itemize}

\item The requirement of a positive definite scalar potential leads to a negligible contribution of the angular mode $\theta$ for inflation (see table~\ref{tab:numdn}), that is $\mu_H \leq 10^{-7}$;

\item The scalar potential for single field inflation takes the form: $V(\varphi) = V_0^{(i)} \left(1-e^{-2\varphi/\sqrt{6}} \right)^2$, with $i=H$, $\Delta$ and ${\rm mixed}$;

\item The constraints from neutrino masses and Higgs mass (see discussion on $\mu_H$ and $M_\Delta$ in the previous section) sets $M_\Delta \lsim {\cal O} (10^8)$ GeV and $\mu_H/M_\Delta = f_H < 0.1$.

\end{itemize}

We see that the inflationary constraints are not capable to put stringent limits for the generation of the CP asymmetries and vice-versa, because different couplings are involved in each step. We therefore discuss separately the inflationary and DM-leptogenesis constraints. The sampling of the parameter space is performed via Markov-Chain Monte Carlo (MCMC) techniques, using a modified version of the public codes \texttt{CosmoMC}~\cite{Lewis:2002ah,cosmomc_notes} and \texttt{SuperBayes}~\cite{superbayes}. In all plots the triplet mass is fixed at $10^8$ GeV and $M_H=125$ GeV.

\subsection{High energy scale -- Single field inflation}
\label{sec:hessf}

\begin{table}[t!]
\caption{MCMC parameters and priors for the scalar potential parameters and non-minimal coupling to gravity at EW scale $\mu=m_t$. All priors are uniform over the indicated range.\label{tab:priorMix}}
\begin{center}
\begin{tabular}{ll}
\hline
 MCMC parameter & Prior \\
\hline
 $\lambda_\Delta$  & $0.1 \to 1$\\ 
  $\lambda_{H\Delta}$ &  $-1 \to  1.2$\\
 $\log_{10}\xi_i$ with $i=H,\Delta$ & $0 \to 6$ \\
 $f_L,f_\psi$ &  $0.005 \to 1$\\
  $\mu_H/M_\Delta$ & $0.009 \to 0.09$ \\
\hline
\end{tabular}
\end{center}
\end{table}

The region of the parameter space compatible with inflation is obtained by solving the RG equations from EW scale up to unitarity scale ($\mu_U \equiv{\rm min}(M_{\rm pl}/\xi_H, M_{\rm pl}/\xi_\Delta)$), and by imposing the constraints on both the power spectrum measured by WMAP7 (\ref{eq:pptheo}), and  the positivity of the scalar potential. The likelihood follows a gaussian distribution centered on the measured value of the primordial density perturbations. Further constraints are:
\begin{enumerate}
	\item All couplings at the unitary scale should satisfy the perturbativity bound: $\lambda_i < \sqrt{4 \pi}$;
	\item All quartic couplings should be definite positive at all scales, the run down to negative values is forbidden: $\lambda_\Delta, \lambda_H > 0$ (below $M_\Delta$, $\lambda_H=\Lambda$);
	\item $\lambda_{H\Delta} +\sqrt{\lambda_H \lambda_\Delta} > 0$;
	\item The vacuum energy should be positive: $V(\varphi) \geq 0$.
\end{enumerate}
The sampling is performed over 7 parameters, listed in table~\ref{tab:priorMix}, together with their uniform prior ranges.

\begin{figure}[t!]
\centering
\includegraphics[width=0.8\columnwidth,trim=0mm 120mm 0mm 65mm, clip]{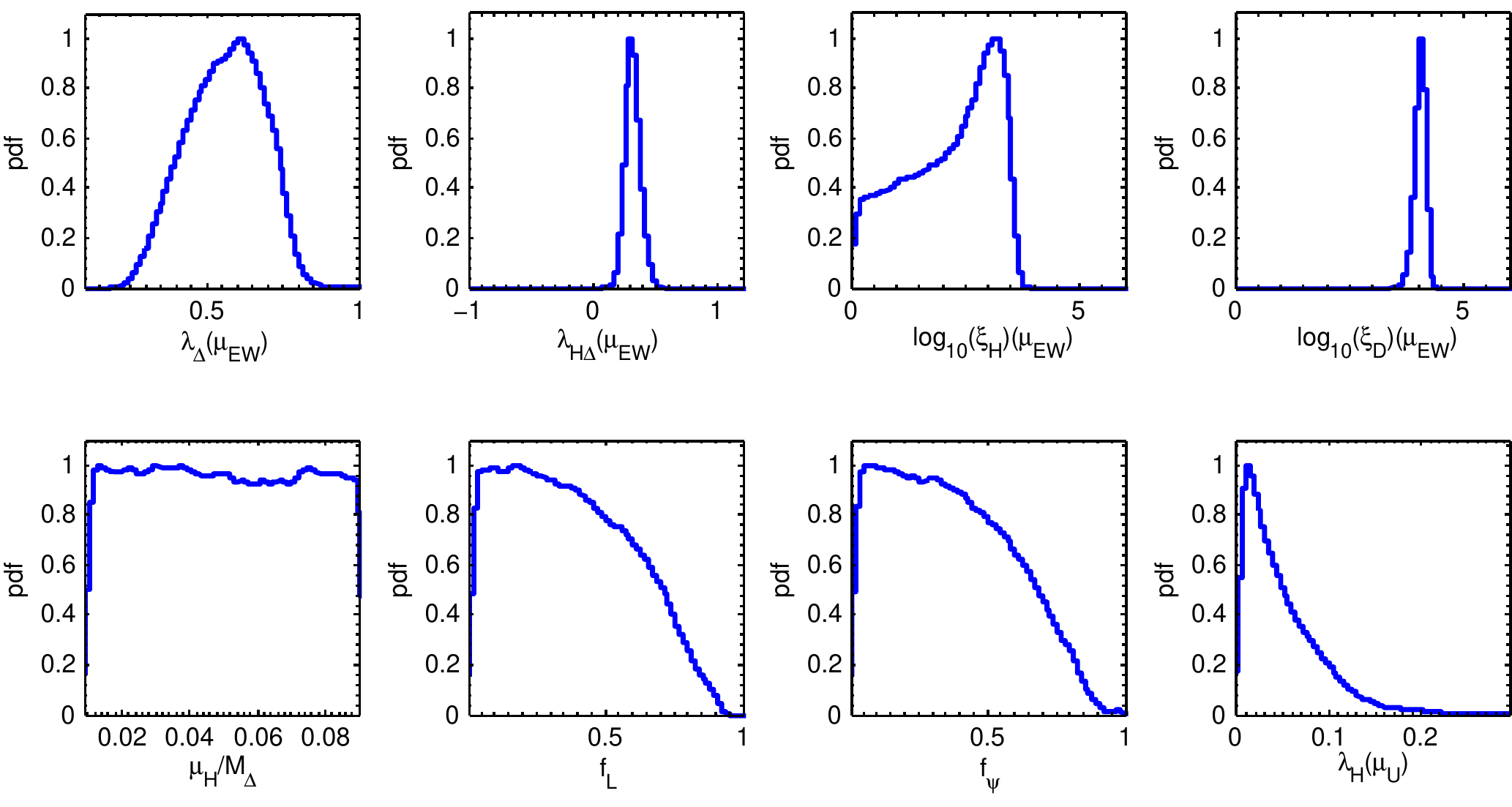}
\caption{1D marginal posterior for $\lambda_\Delta$,  $\lambda_{H\Delta}$, $\log_{10}\xi_i$ (with $i=H$, $\Delta$), $\mu_H/M_\Delta$, $f_L$ and $f_\psi$ at EW scale. In addition we show the 1D marginal posterior for $\lambda_H$ at the unitarity scale $\mu_U$, while at EW scale $\lambda_H=0.26$. All of other parameters in each plane have been marginalized over.}
\label{fig:mixsf_1D}
\end{figure}

\begin{figure}[t!]
\begin{minipage}[t]{0.33\textwidth}
\centering
\includegraphics[width=1.\columnwidth,trim=40mm 90mm 50mm 90mm,clip]{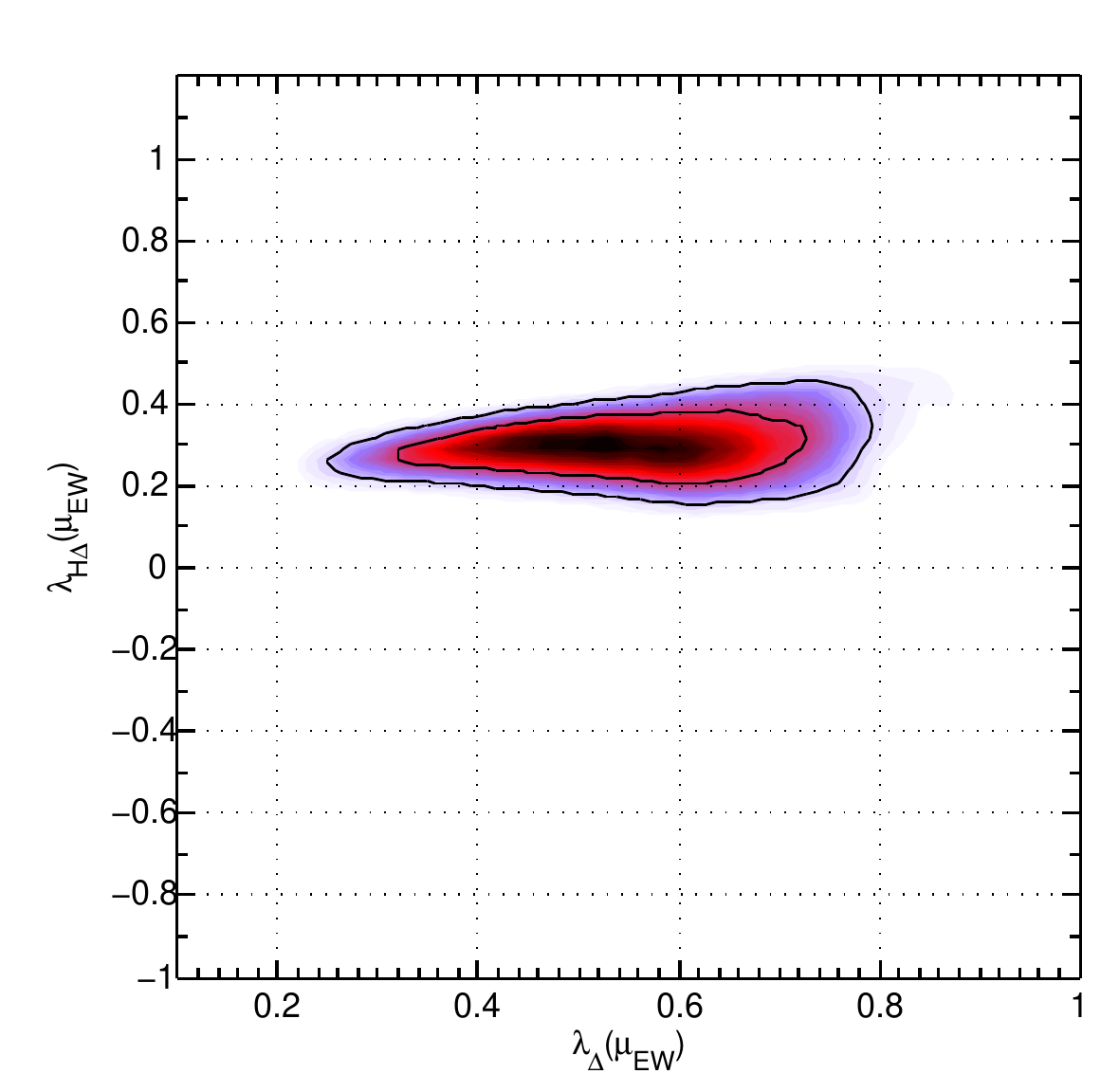}
\end{minipage}
\hspace*{-0.2cm}
\begin{minipage}[t]{0.33\textwidth}
\centering
\includegraphics[width=1.\columnwidth,trim=40mm 90mm 50mm 90mm,clip]{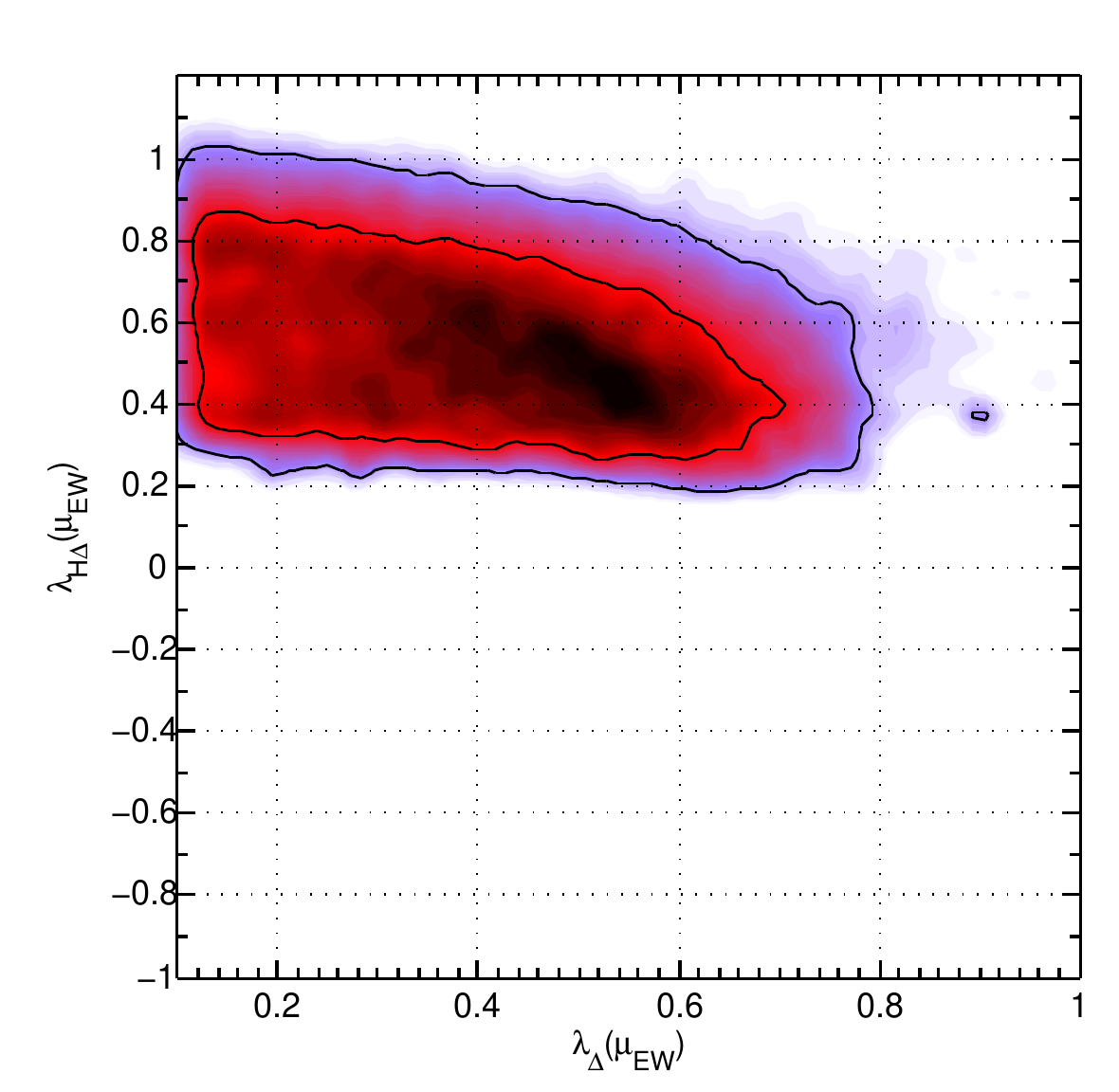}
\end{minipage}
\hspace*{-0.2cm}
\begin{minipage}[t]{0.33\textwidth}
\centering
\includegraphics[width=1.\columnwidth,trim=13mm 25mm 15mm 15mm,clip]{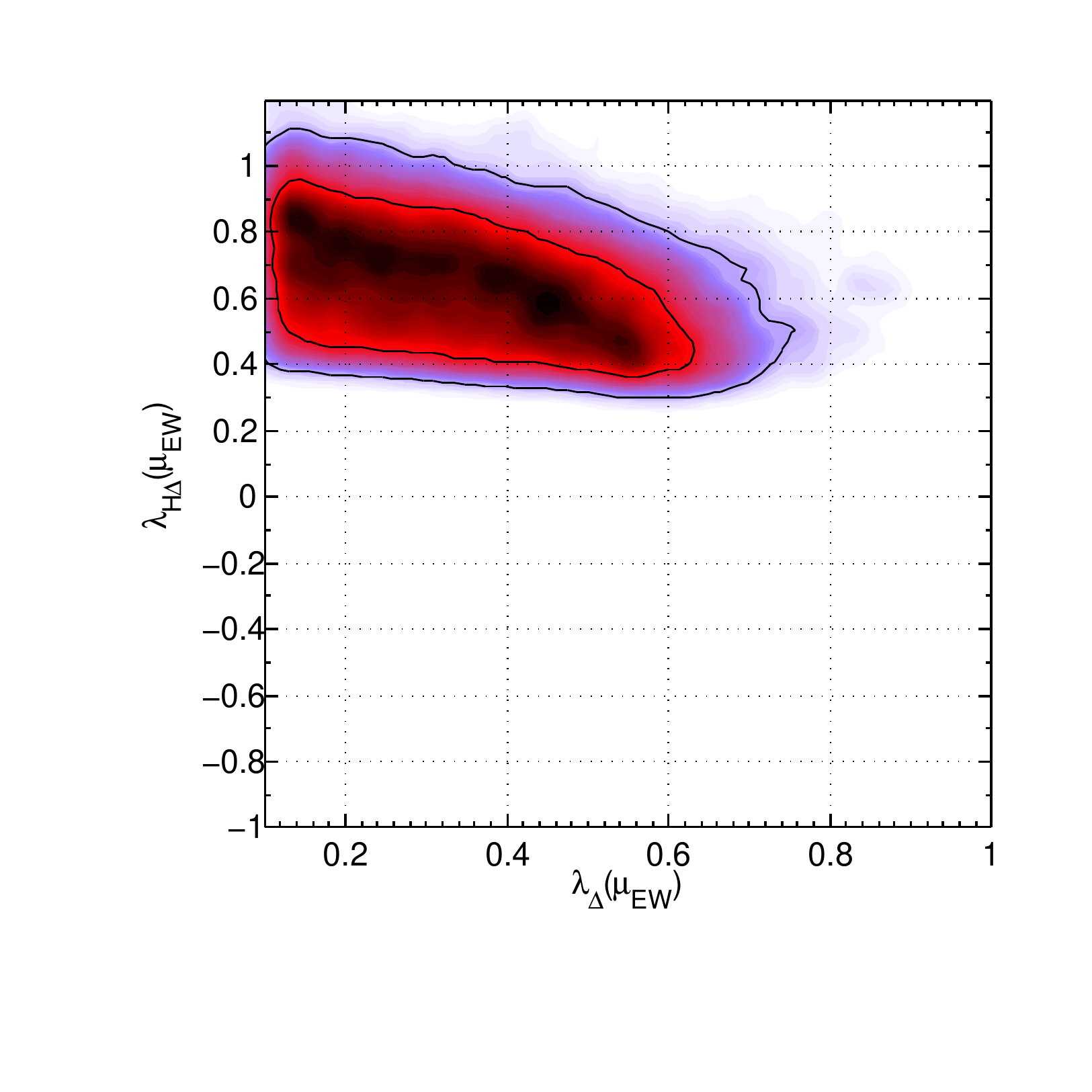}
\end{minipage}
\caption{{\it Left:} 2D marginal posterior in the $\{\lambda_{\Delta}(\mu_{\rm EW}),\lambda_{H\Delta}(\mu_{\rm EW})\}$-plane for mixed inflation. The black solid lines enclose the 68\% and 95\% credible region. {\it Center and right:} Same as left for pure Higgs and pure triplet inflation respectively. All of other parameters in each plane have been marginalized over.}
\label{fig:2Dlhdld}
\end{figure}

\paragraph{Mixed inflaton $\equiv V_0^{(\rm mixed)}$}
We require additionally the conditions (\ref{eq:constMIx1}), (\ref{eq:constMIx2}) and (\ref{eq:constMIx3}) should be satisfied, in order $r=r_0$ to be a positive minimum of the potential $V(r)$, given equation (\ref{eq:vrpot}). Figure~\ref{fig:mixsf_1D} shows the 1D marginal posterior probability distribution functions (pdf) for all the MCMC parameters at EW scale plus the results for $\lambda_H$ at the unitarity scale. The preferred values for $\lambda_\Delta$ span all the sampled range with a preference for the central values from $0.2$ up to $0.8$, while $\lambda_{H\Delta}$ is very constrained. This is a result of (\ref{eq:constMIx1}). The 2D credible region in the plane $\{\lambda_{\Delta}(\mu_{\rm EW}),\lambda_{H\Delta}(\mu_{\rm EW})\}$ is shown in the left panel of figure~\ref{fig:2Dlhdld}, which ultimately indicates that $\lambda_{H\Delta}$ is constrained to be in the range $0.2 \to 0.4$ and positive definite. Even though the non-minimal couplings to gravity are not observable, we show their preferred value in the third and fourth panels of the first raw of figure~\ref{fig:mixsf_1D}: $\xi_\Delta$ follows a distribution sharply peaked around $10^4$, while $\xi_H$ prefers smaller values than $\xi_\Delta$ in a much broader range: from 1 up to $10^4$. Those values prevent the unitarity scale to be too low and insure the requirement that the quartic term dominates the potential. The parameter for lepton violation (second raw, first panel) is loosely constrained: no value is found to be significantly preferred. Same for the Yukawa couplings $f_L$ and $f_\psi$ (second and third panels, second raw), except that values close to one are slightly disfavored. Eventually the last panel shows $\lambda_H(\mu_U)$, which is positive definite of course and can reach at most values of about $0.3$. 

\begin{figure}[t!]
\centering
\includegraphics[width=0.8\columnwidth,trim=0mm 120mm 0mm 65mm, clip]{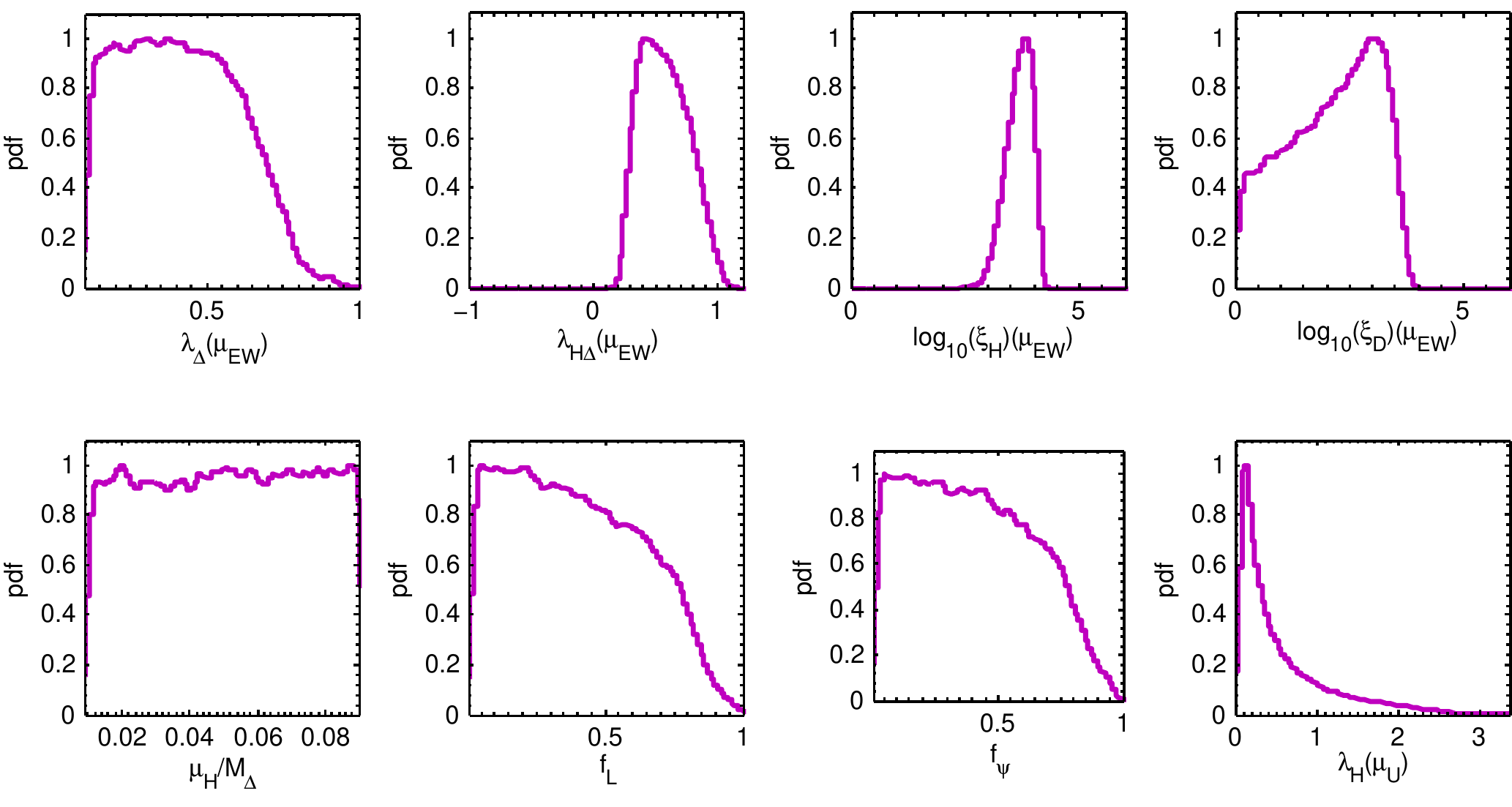}
\caption{Same as figure~\ref{fig:mixsf_1D} for pure Higgs inflation.}
\label{fig:hsf_1D}
\end{figure}

\paragraph{Pure Higgs as inflaton $\equiv V_0^{(H)}$}
We require additionally the conditions (\ref{pureHiggs_condition1}) and (\ref{pureHiggs_condition2}) should be satisfied for the positivity and stability of the vacuum energy. The 1D marginalised posterior pdfs are shown in figure~\ref{fig:hsf_1D}: the first and second panel denote the preferred values for $\lambda_\Delta$ and $\lambda_{H\Delta}$. These couplings are less constrained than the mixed case, since $\lambda_{H\Delta}$ can acquire any positive values from 0.2 up to 1, although the pdf is peaked for values of about 0.5-0.6. The 2D credible region as a function of both couplings is plotted in figure~\ref{fig:2Dlhdld}, middle panel: note that it is larger than the mixed case. The non-minimal coupling of the Higgs to gravity is well constrained and its pdf is peaked at $10^4$, while $\xi_\Delta$ can span a broad range of values from 1 up to $10^4$. The parameters related to DM and leptons are unconstrained as in the previous case. The Higgs quartic coupling can reach large values up to perturbativity bound. Similar pdfs for all the couplings are obtained setting $\xi_\Delta$ to zero.

\begin{figure}[t!]
\centering
\includegraphics[width=.8\columnwidth,trim=0mm 120mm 0mm 65mm, clip]{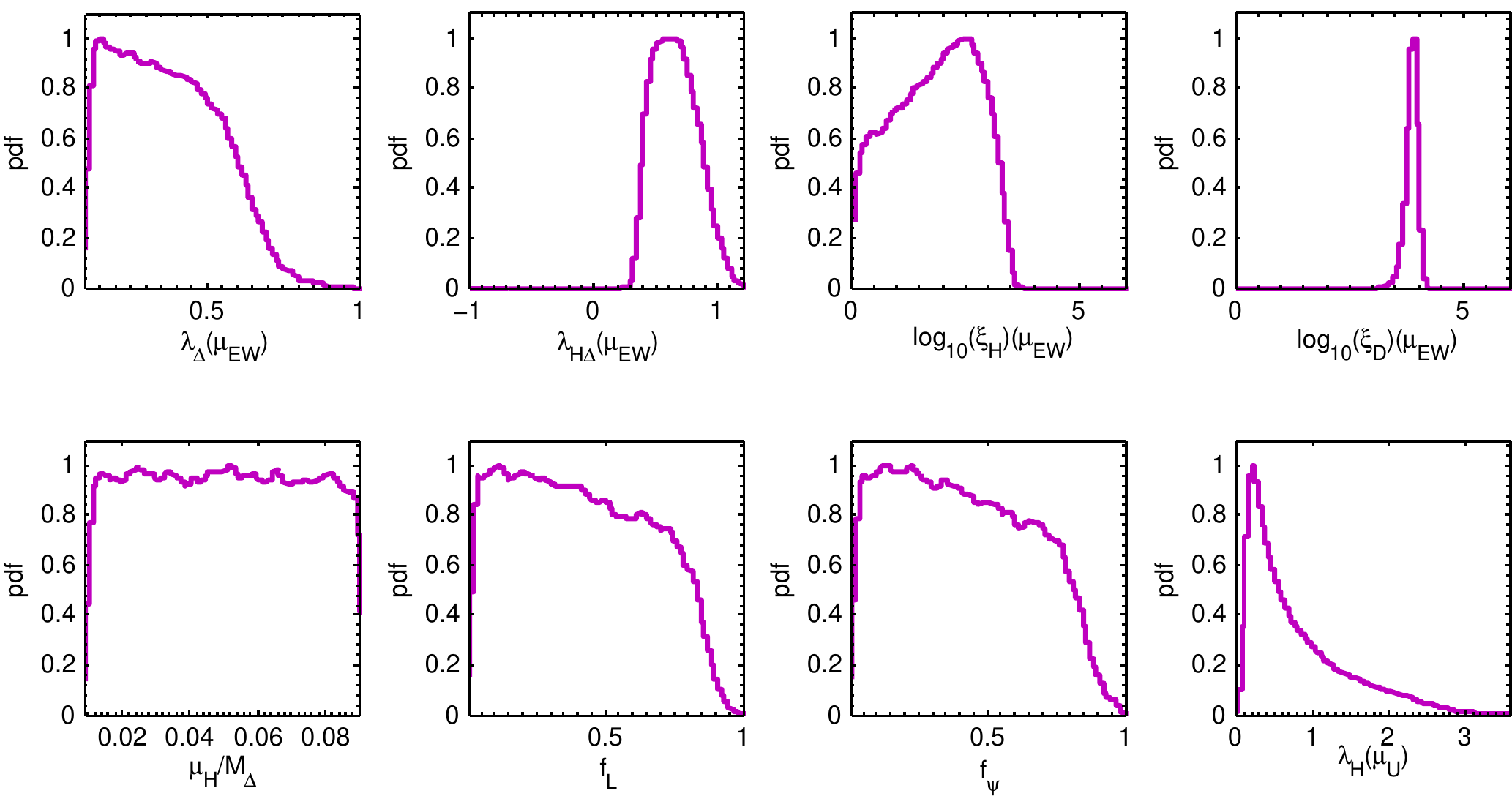}
\caption{Same as figure~\ref{fig:mixsf_1D} for pure triplet inflation.}
\label{fig:tsf_1D}
\end{figure}

\paragraph{Pure triplet as inflaton $\equiv V_0^{(\Delta)}$}
Here we require additionally the conditions (\ref{puretriplet_condition1}) and (\ref{puretriplet_condition2}) should be satisfied. For all parameters but $\xi_H$ and $\xi_\Delta$, the 1D marginal posterior pdfs are very similar to the case pure Higgs inflation (figure~\ref{fig:tsf_1D}) as well as the 2D credible region for $\lambda_\Delta$, $\lambda_{H\Delta}$ (figure~\ref{fig:2Dlhdld}, right panel). From figure~\ref{fig:tsf_1D}, the third and fourth panel (first raw) depict the behavior of the non-minimal couplings to gravity. $\xi_H$ is essentially unconstrained and can vary with almost equal probability from 1 up to $10^4$, while $\xi_\Delta$ is described by a narrow gaussian centered on its mean value $\mathcal{O}(10^4)$, a case similar to mixed inflation. Note that the case of $\xi_H=0$ gives equivalent results.

\subsection{Low energy scale -- DM and visible sectors}
\label{sec:les}

The abundances of matter in the dark and visible sectors satisfying (\ref{eq:IMP}) are produced by the quasi-equilibrium decay of the triplet scalar $\zeta_1$. The free parameters are the CP asymmetries $\epsilon_i$ for all the species, the branching ratios $B_i$ and the dark matter mass $m_{\rm DM}$. The following constraints apply:
\begin{align}\label{eq:const}
\sum_j \epsilon_j = & \, \, 0\,,  
\\
\sum_j B_j = &  \, \, 1 \,,
\\
 | \epsilon_j| \le  &  \, \, 2 \ B_j\,.
\end{align}
The first and third conditions ensure that all amplitudes are physical and the total amount of CP violation can not exceed 100\% in each channel, while the second one simply demands unitarity of the model. The five free parameters are $\epsilon_L, \epsilon_{\rm DM}, B_L, B_{\rm DM}$ and $m_{\rm DM}$. In addition to that we allow a hierarchy between the CP asymmetries: $\epsilon_H \simeq \epsilon_{\rm DM} \sim 10^{2}-10^{5} \epsilon_{L}$, as remarked in section~\ref{sec:ADML}. From RG evolution and vacuum stability above $10^8$ GeV we require $f_H < 0.1$. Those are the main novelties with respect to the analysis in~\cite{Arina:2011cu}.

The sampling of the parameter space is done via MCMC methods and the parameter inference is Bayesian, following the same setup of~\cite{Arina:2011cu}. We note that the likelihood is given by the sum of a ratio function satisfying~(\ref{eq:IMP}), that is $r \equiv \Omega_{\rm DM}/\Omega_b$, plus a gaussian distribution describing the baryon to photon ratio, centered on $\bar{\eta}_b \pm \sigma_{\eta b} = (6.15 \pm 0.25) \times 10^{-10}$. The prior ranges are given in table~\ref{tab:priorDM}. The DM allowed range starts at 45 GeV: doublet candidates are excluded below this value by the invisible decay width of the $Z$ boson.
\begin{table}[t!]
\caption{MCMC parameters and priors for the CP asymmetries, branching ratios and $m_{\rm DM}$. All priors are uniform over the indicated range.\label{tab:priorDM}}
\begin{center}
\begin{tabular}{ll}
\hline
 MCMC parameter & Prior \\
\hline
 $\log_{10}(m_{\rm DM}/{\rm GeV})$  & $1.66  \to 3$\\ 
  $\log_{10}\epsilon_{\rm DM}$ &  $-5 \to  -2$\\
 $\log_{10}\epsilon_L$ & $-9 \to -4$ \\
 $\log_{10}B_{\rm DM}$ &  $-2 \to 0$\\
  $\log_{10}B_L$ & $-4.5 \to -3$ \\
\hline
\end{tabular}
\end{center}
\end{table}
\begin{figure}[t]
\begin{minipage}[t]{0.5\textwidth}
\centering
\includegraphics[width=1.\columnwidth,trim=6mm 5mm 15mm 30mm, clip]{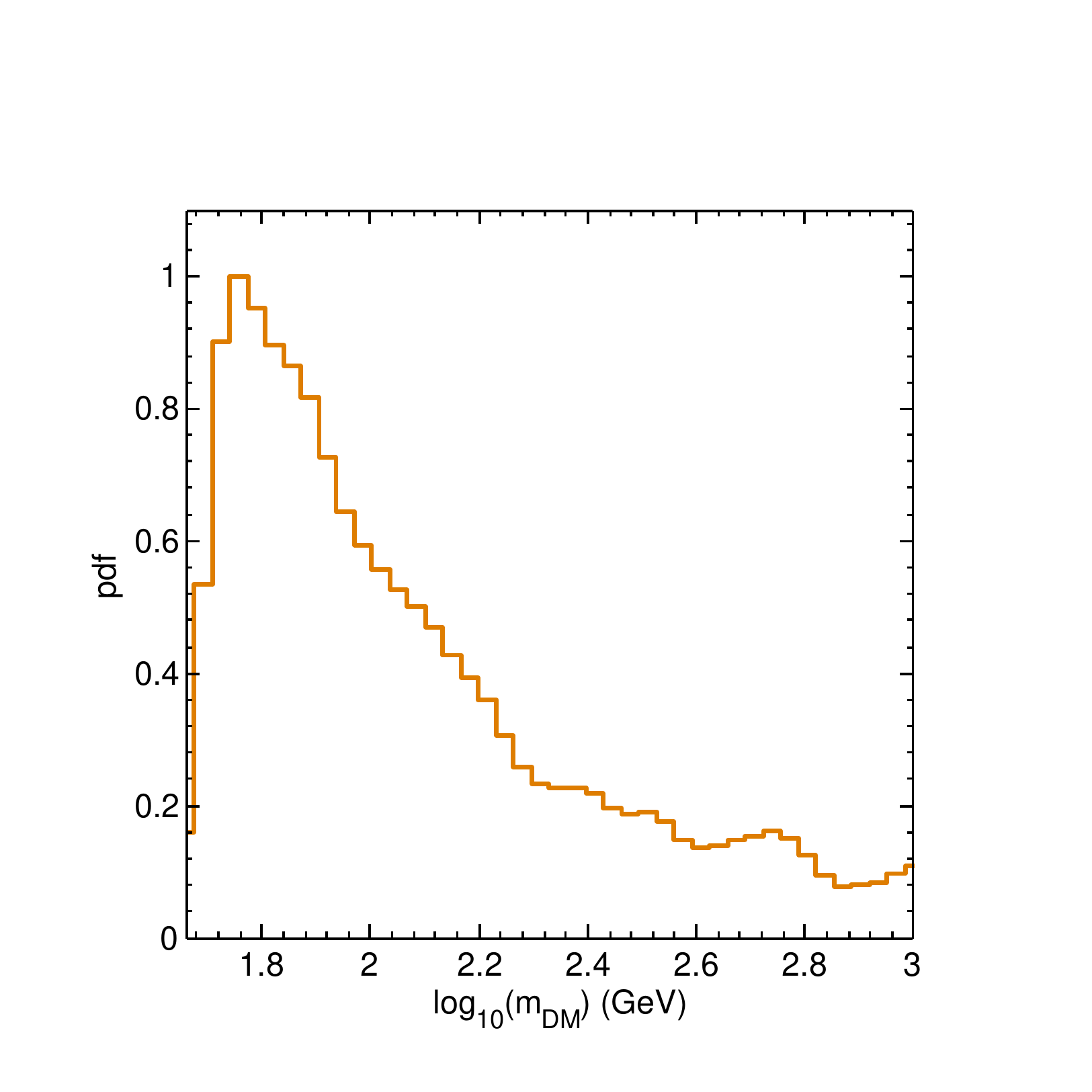}
\end{minipage}
\hspace*{-0.2cm}
\begin{minipage}[t]{0.5\textwidth}
\centering
\includegraphics[width=1.\columnwidth,trim=15mm 25mm 10mm 15mm, clip]{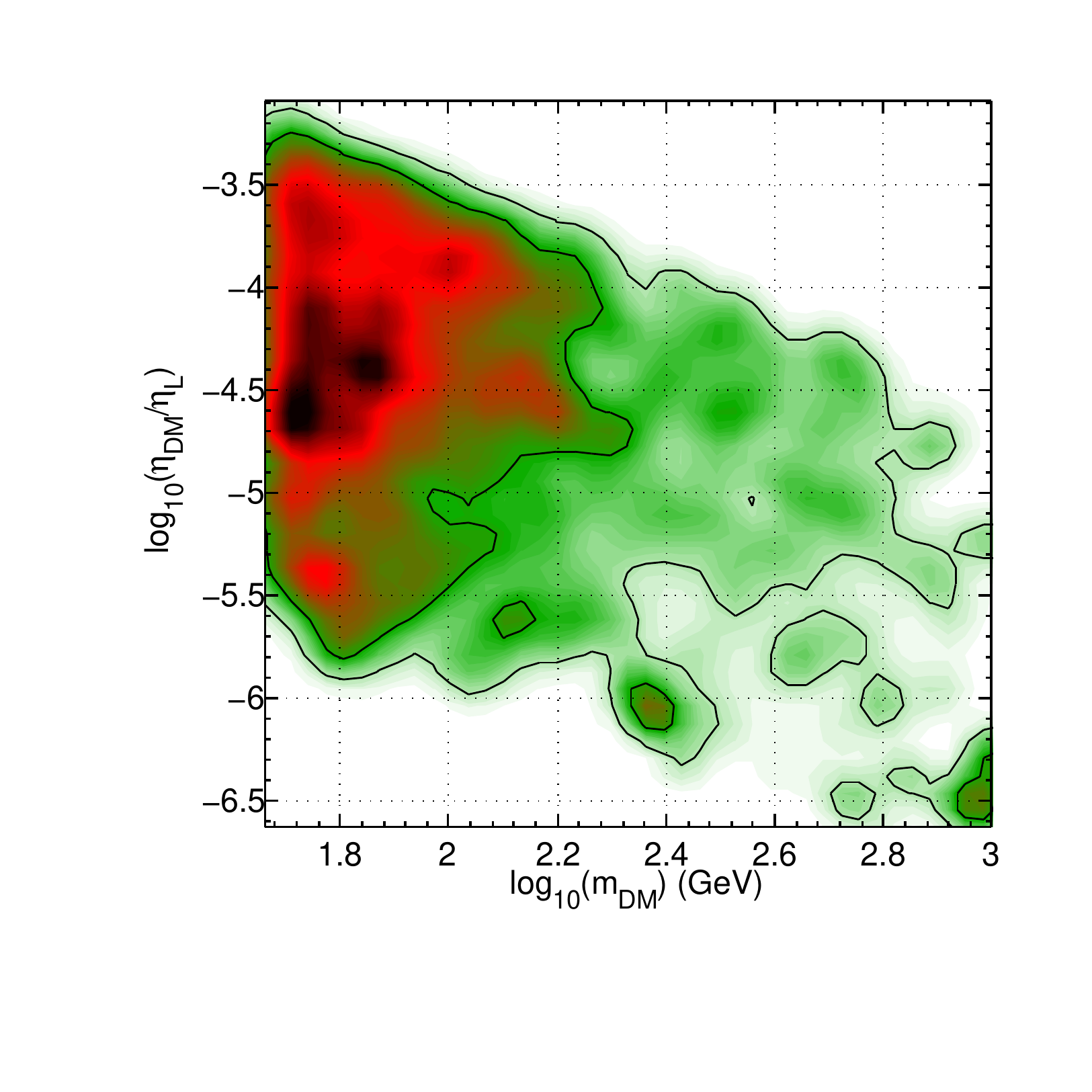}
\end{minipage}
\caption{{\it Left:} 1D posterior pdf for the DM mass $m_{\rm DM}$. {\it Right:} 2D credible regions at 68\% and 95\% C.L. in the \{$m_{\rm DM},\eta_{\rm DM}/\eta_{L}$\}-plane. All of other parameters in each plane have been marginalized over.}
\label{fig:1DmDM}
\end{figure}
\begin{figure}[t!]
\begin{minipage}[t]{0.5\textwidth}
\centering
\includegraphics[width=1.\columnwidth,trim=15mm 25mm 10mm 15mm, clip]{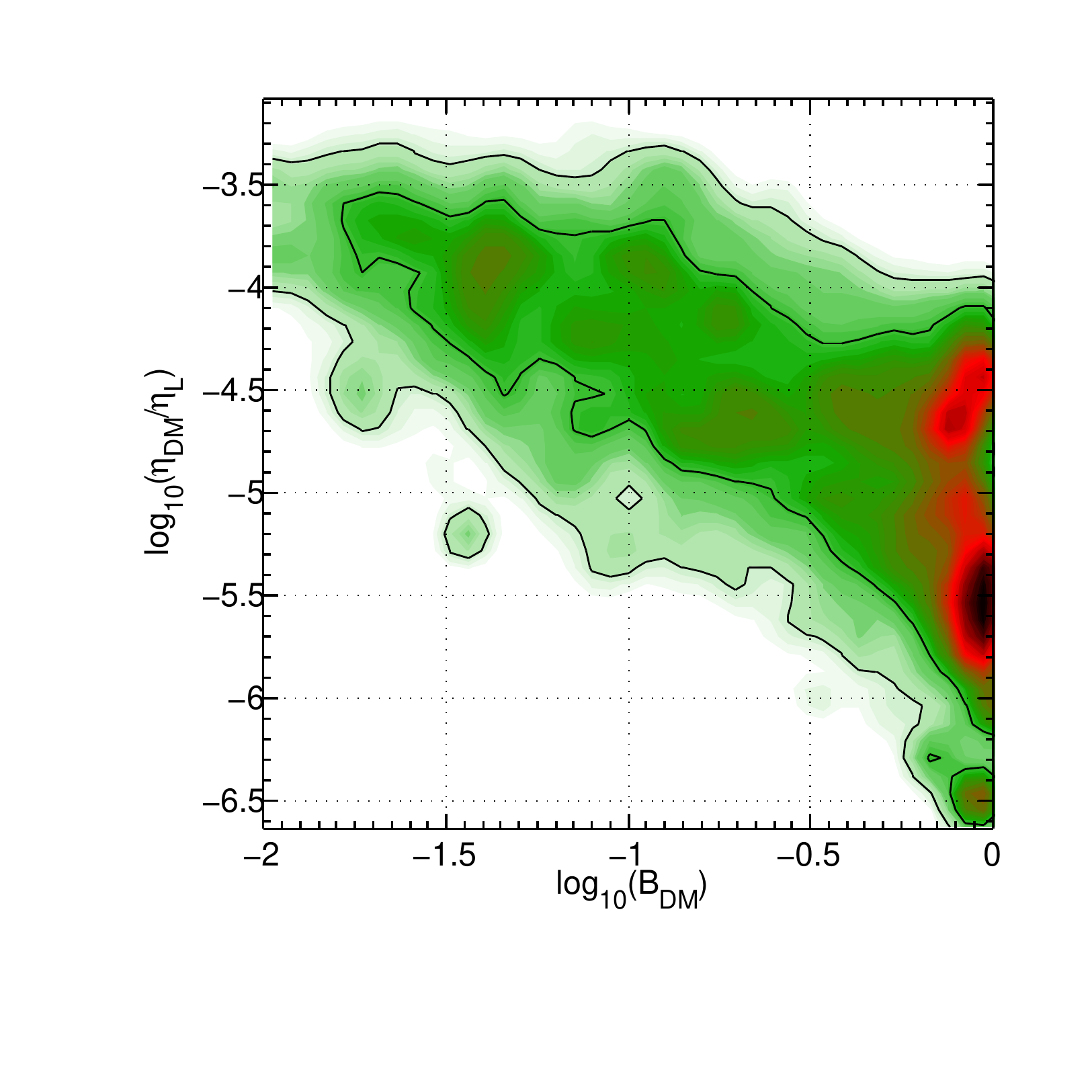}
\end{minipage}
\hspace*{-0.2cm}
\begin{minipage}[t]{0.5\textwidth}
\centering
\includegraphics[width=1.\columnwidth,trim=15mm 25mm 10mm 15mm, clip]{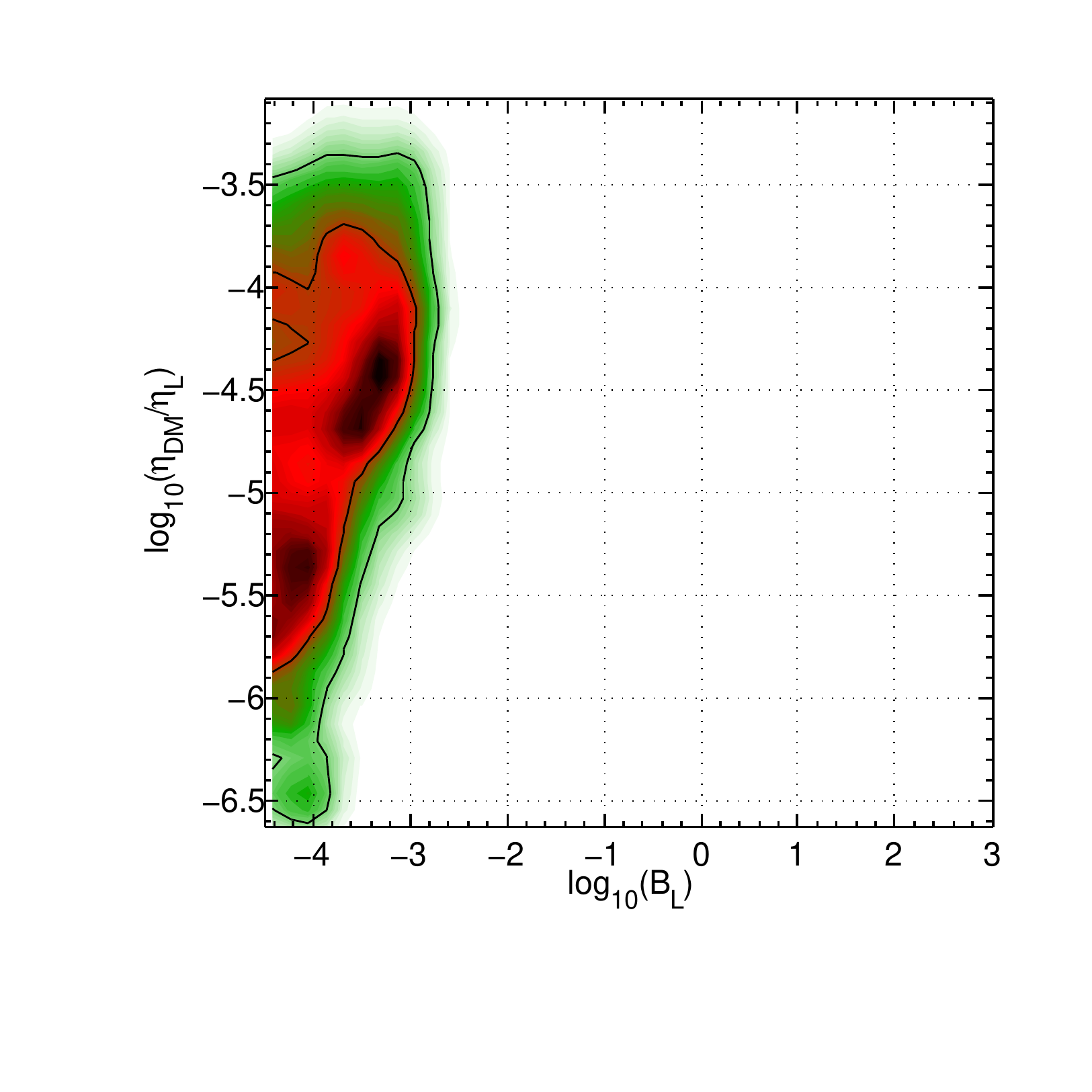}
\end{minipage}
\caption{ {\it Left:} 2D posterior pdf in the \{$B_{\rm DM},\eta_{\rm DM}/\eta_{L}$\}-plane. {\it Right:} Same as left in the \{$B_{\rm L},\eta_{\rm DM}/\eta_L$\}-plane. The credible regions are given at $68\%$ and $95\%$ C.L. All of other parameters in each plane have been marginalized over.
\label{fig:set2}}
\end{figure}
\begin{figure}[t!]
\begin{minipage}[t]{0.33\textwidth}
\centering
\includegraphics[width=1.\columnwidth,trim=15mm 25mm 10mm 15mm, clip]{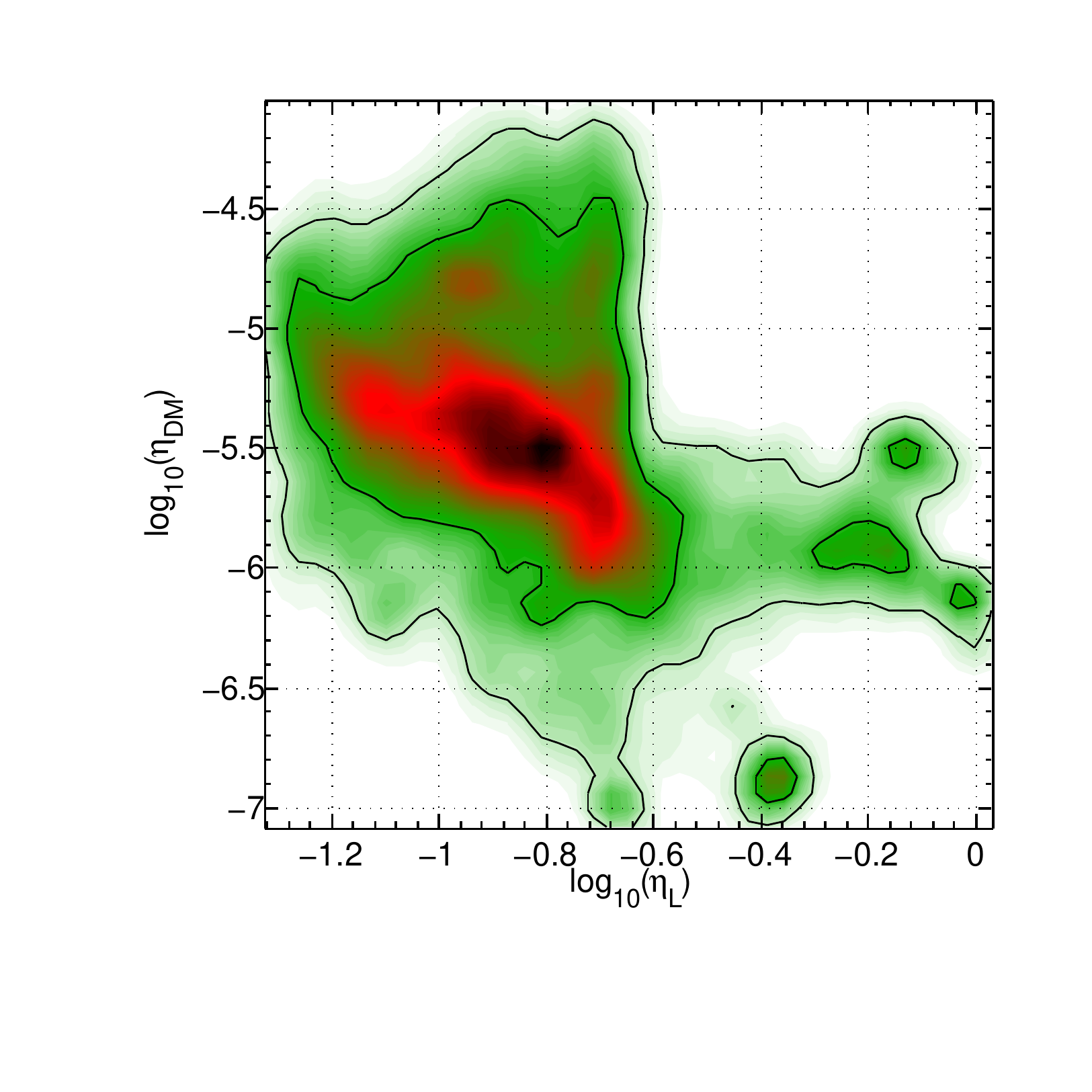}
\end{minipage}
\begin{minipage}[t]{0.33\textwidth}
\centering
\includegraphics[width=1.\columnwidth,trim=15mm 25mm 10mm 15mm, clip]{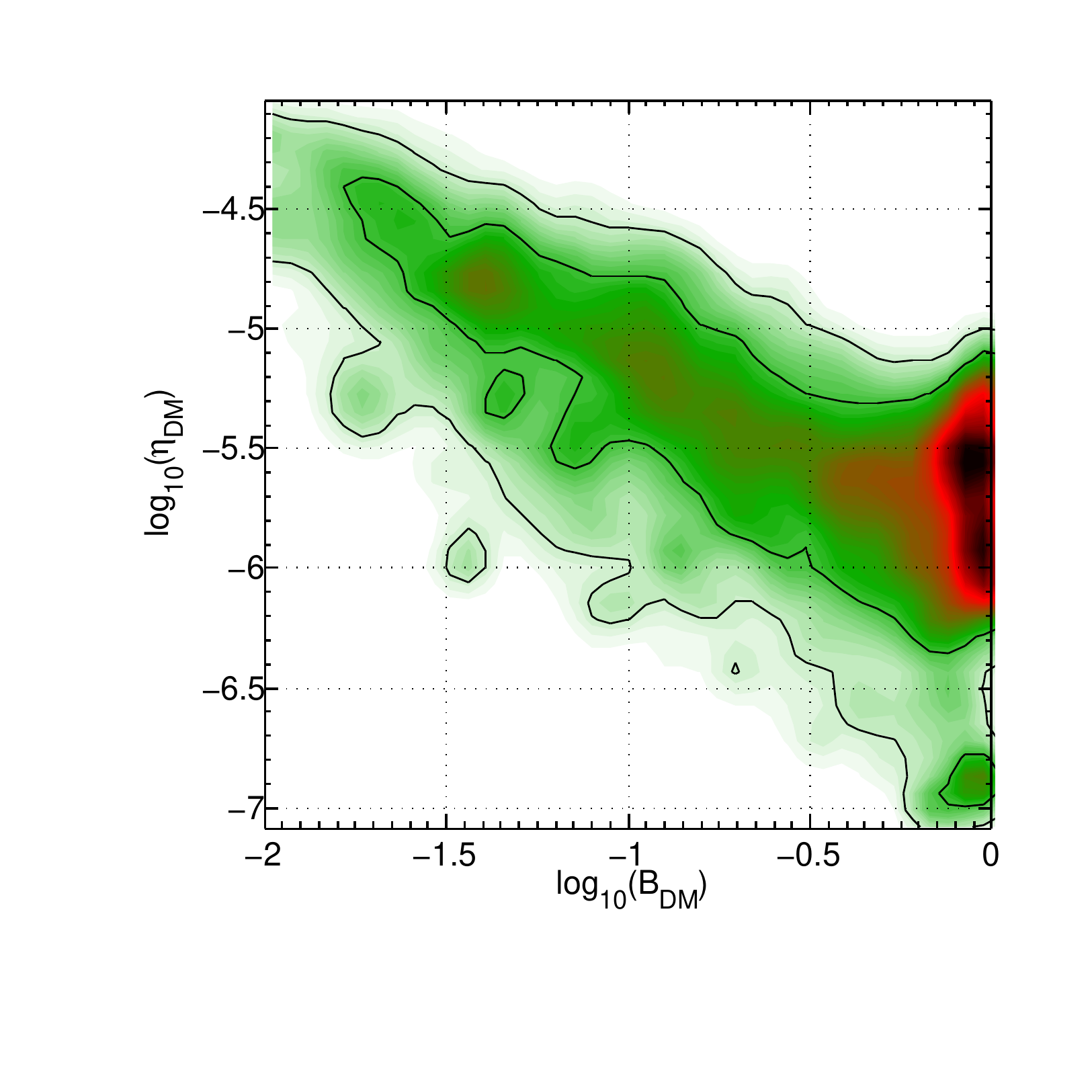}
\end{minipage}
\hspace*{-0.2cm}
\begin{minipage}[t]{0.33\textwidth}
\centering
\includegraphics[width=1.\columnwidth,trim=15mm 25mm 10mm 15mm, clip]{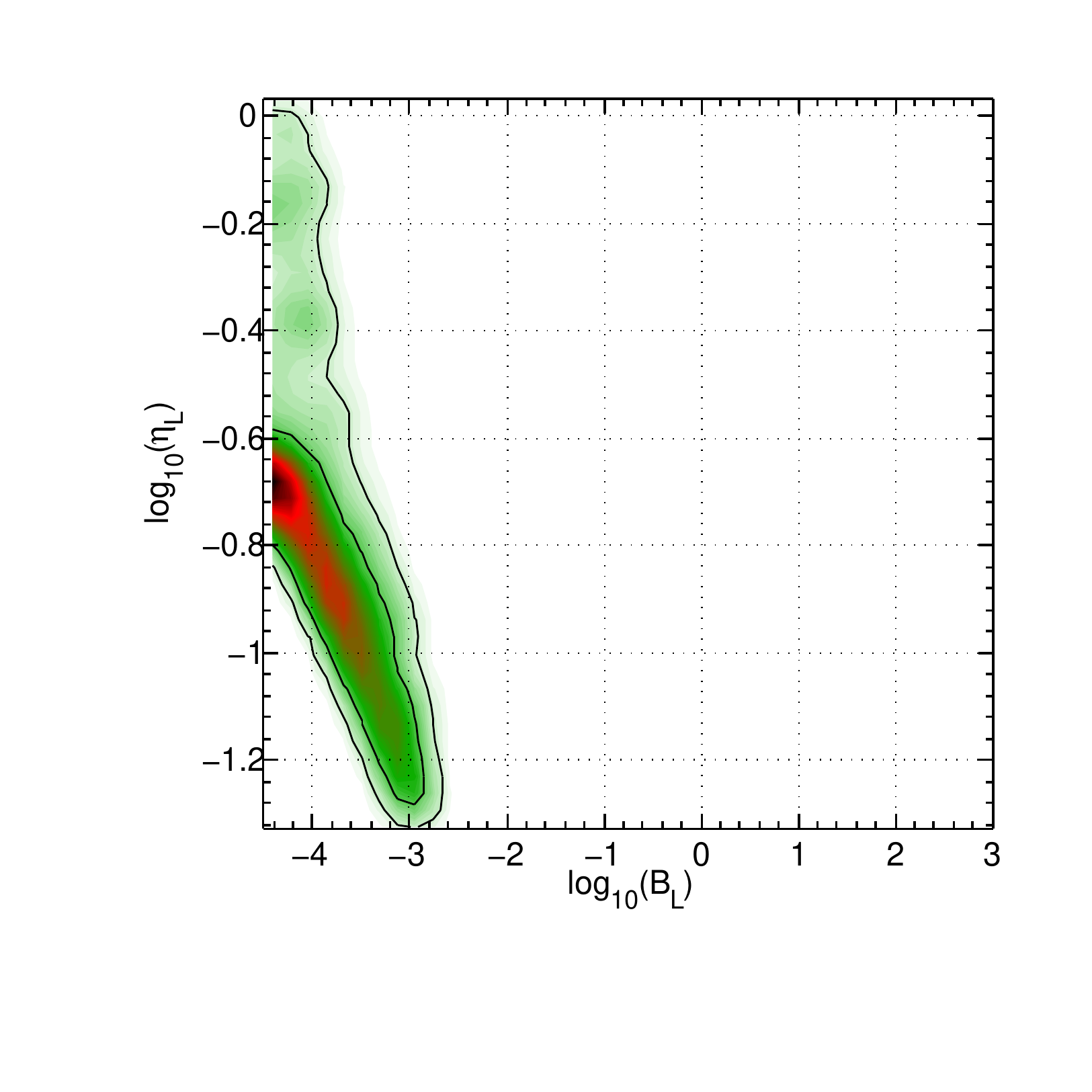}
\end{minipage}
\caption{{\it Left:} 2D posterior pdf in the \{$\eta_{\rm DM},\eta_{L}$\}-plane. {\it Central:}  2D posterior pdf in the \{$B_{\rm DM},\eta_{\rm DM}$\}-plane. {\it Right:} same as left in the \{$B_{L},\eta_{L}$\}-plane. The credible regions are given at $68\%$ and $95\%$ C.L. All of other parameters in each plane have been marginalized over.
\label{fig:set3}}
\end{figure}
\begin{figure}[t]
\begin{minipage}[t]{0.5\textwidth}
\centering
\includegraphics[width=1.\columnwidth,trim=20mm 80mm 35mm 90mm, clip]{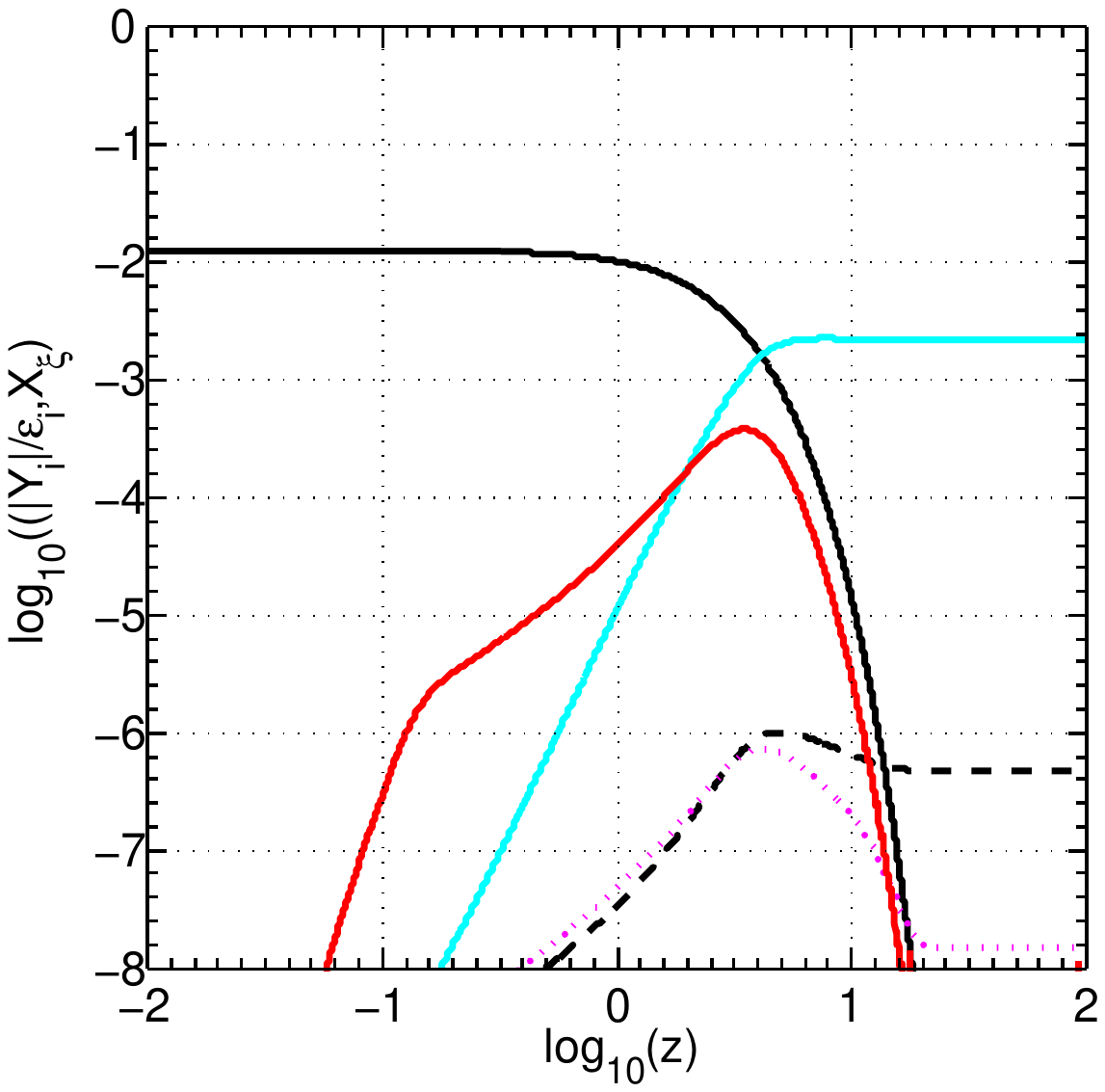}
\end{minipage}
\hspace*{-0.2cm}
\begin{minipage}[t]{0.5\textwidth}
\centering
\includegraphics[width=1.\columnwidth,,trim=20mm 80mm 35mm 90mm, clip]{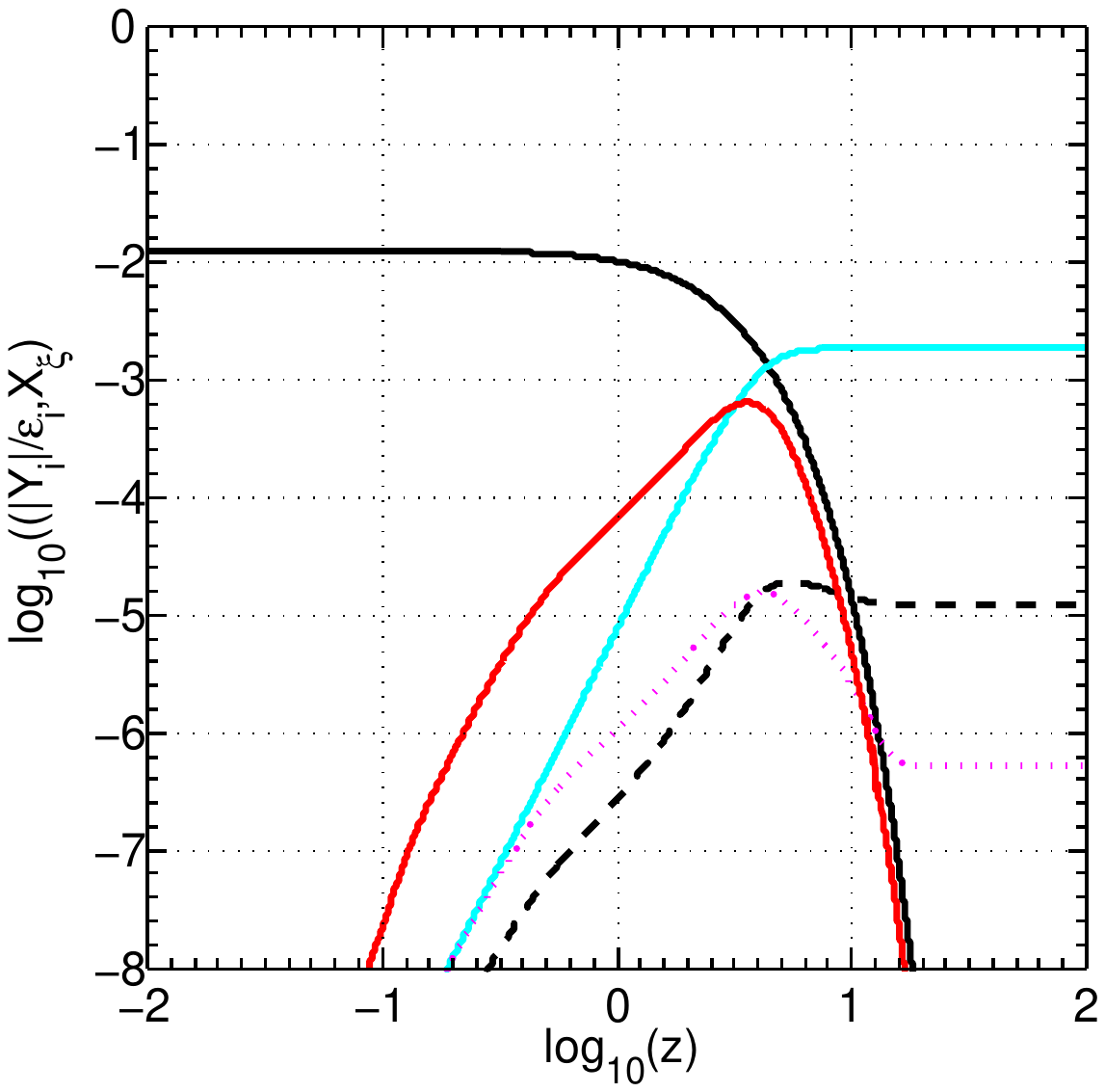}
\end{minipage}
\caption{Absolute value for the Yield of leptons (cyan solid), DM (dotted magenta), Higgs (dashed black), $\zeta$ asymmetry (solid red) plus scalar triplet abundancy (black solid) for two successful points as follows. {\it Left:}  $m_{\rm  DM}=76$ GeV, $B_L=9.1\times 10^{-5}$, $B_{\rm DM}=0.51$, $\epsilon_{L}=6.8 \times 10^{-8}$, $\epsilon_{\rm DM}=3.3\times  10^{-4}$ which leads to $r\equiv \Omega_{\rm DM}/\Omega_b=4.7$, $Y_L=1.53 \times 10^{-10}$ and $\eta_{\rm DM}/\eta_L=6.49 \times 10^{-6}$. {\it Right:} $m_{\rm DM}=60$ GeV, $B_L=8.6 \times 10^{-5}$, $B_{\rm DM}=0.017$, $\epsilon_L=8.9 \times 10^{-8}$, $\epsilon_{\rm DM}=1.4 \times 10^{-5}$, $\Omega_{\rm DM}/\Omega_b=5.1$ and $Y_L=1.64\times 10^{-10}$ and $\eta_{\rm DM}/\eta_L=2.8 \times 10^{-4}$. The $|Y_i|$ are rescaled in terms of CP asymmetries.}
\label{fig:points}
\end{figure}

In the left panel of figure~\ref{fig:1DmDM} we show the 1D posterior pdf for $m_{\rm DM}$, while all other parameters are marginalized over. We see that all the mass range from 45 GeV up to 1 TeV can lead to successful leptogenesis, namely $Y_L \sim 10^{-10}$ and an asymmetric dark matter candidate satisfying (\ref{eq:IMP}). Note from the posterior pdf that the most favored region is for low mass candidates, even though there are candidates viable up to 1 TeV with smaller statistical significance. On the right panel of figure~\ref{fig:1DmDM} the 68\% and 95\% credible regions are shown in the \{$m_{\rm DM},\eta_{\rm DM}/\eta_{L}$\}-plane. From there we see that for DM mass up to around 150 GeV, the preferred values of the ratio $\eta_{\rm DM}/\eta_L$ are of  $\calO(10^{-4}-10^{-5})$, which compensate the large CP asymmetry ratio: $\epsilon_{\rm DM}/\epsilon_L$. However, for DM masses around ${\cal O}$(TeV), $\epsilon_{\rm DM}/\epsilon_L$ is even larger for the preferred values of $\eta_{\rm DM}/\eta_L$, which decreases down to $10^{-6}$. For a triplet mass of $10^8$ GeV the important quantities which drive Boltzmann equations are the branching ratios. In figure~\ref{fig:set2} we show the correlation of $\eta_{\rm DM}/\eta_L$ versus $B_{\rm DM}$ and $B_L$ respectively in the left and right panels, within the 68\% and 95 \% credible regions. We see that the largest efficiency ratio $\eta_{\rm DM}/\eta_L$ is preferred when $B_{\rm DM} \to 1$ and small $B_{\rm L} \to 10^{-3} - 10^{-4}$. This is because of the required hierarchy between the sub-eV neutrino mass and the Majorana mass splitting between the DM mass eigenstates. We note that without this constraint the preferred values would be the opposite ones, as shown in~\cite{Arina:2011cu}. Since $B_{\rm DM}$ is large, which implies small $B_L$ as $\sum_i B_i=1$ (with i=$L,H$,DM), the washout is large as well, which leads to small $\eta_{\rm DM}$. On the contrary smaller is the $B_L$ the washout effect is small due to inverse decay and hence large $\eta_L$ is preferred. Note that in either case the production of asymmetry is proportional to $\Gamma_1 \propto 1/\sqrt{B_L B_H}$. Therefore when $B_L$ approaches towards $10^{-5}$ the asymmetry ($Y_L$) as well as the efficiency ($\eta_L$) get increased. On the other hand when $B_{\rm DM}$ approaches towards $1$, which implies small $B_L$, the asymmetry $Y_{\rm DM}$ gets increased but efficiency gets decreased. These behaviors of $\eta_{\rm DM}$ and $\eta_L$ can be confirmed from figure~\ref{fig:set3} where we have shown the 2D credible regions at 68\% and 95\% C.L. The extreme left one, which constitutes the summary of middle and right ones, reveals that a successful asymmetric dark matter and lepton asymmetry can be generated with small $\eta_{\rm DM}$ and large $\eta_L$.  For large $B_{\rm DM}$ and small $B_L$, the CP asymmetry in the DM sector should be larger by more than an order of magnitude with respect to $\epsilon_L$ to compensate the small value of $\eta_{\rm DM}/\eta_L$. The same behavior is recovered for large DM masses.

To illustrate the mechanism of the generation of the asymmetries via triplet decay we show two benchmark points in figure~\ref{fig:points}, which are representative examples from our sampling. In all cases the slow channel that builds and conserves the asymmetry is the leptonic one: the smallness of its branching ratio is due to the hierarchy between neutrino and DM Majorana masses. The fast channels are both the Higgs, to compensate the neutrino mass in $\Gamma_1$, and the DM one. Since the DM channel is not related to the neutrino mass via $\Gamma_1$, its branching ratio can assume different values all along the DM mass range. The first point in the parameter space is shown in the left panel, which leads to a successful model for the DM with a mass of $\sim 76$ GeV, $r \sim 4.7$ and $Y_L=1.5 \times10^{-10}$. The second point in the parameter space is depicted in the right panel and accounts for a DM candidate with $m_{\rm DM}\sim 60$ GeV, $r \sim 5.1$ and successful baryon asymmetry $Y_L= 1.6 \times 10^{-10}$. The details about the parameters are given in the caption. For the left panel the branching ratios are $B_L= 9 \times 10^{-5}$ and $B_{\rm DM}=0.51$, which implies large $\eta_L$ and small $\eta_{\rm DM}$. Therefore, the ratio of $\eta_{\rm DM}/\eta_L \simeq 6 \times 10^{-6}$ is small and can be confirmed from figure~\ref{fig:set2}. The fastest channel will be the DM one, with the largest branching ratio. For the figure in right panel the branching ratio for leptons is comparable with the other benchmark, $B_L=8.5 \times 10^{-5}$, while the DM one, $B_{\rm DM}=0.017$, is much smaller. This implies a larger value for the ratio $\eta_{\rm DM}$ and $\eta_L$, because in this case the fastest channel is the Higgs one. The small values for the efficiency ratio are compensated by the large CP asymmetry ratio, as already discussed and confirmed from figure~\ref{fig:set2}. 

The DM symmetric component is depleted by the efficient gauge interactions before it freezes out and is totally negligible at present day~\cite{Chun:2011cc}, while the asymmetric DM abundance in accordance with WMAP is proportional to the Yield $Y_{\rm DM}$.

Thus we see that in a large portion of the parameter space, in particular around DM masses of $\mathcal{O}(100)$ GeV the constraints of having sub-eV neutrino masses and keV mass splitting for mass eigenstates of $\psi$ are satisfied. In this case, the ratio of the CP asymmetries ranges from $10^3$ up to $10^5$ and (\ref{CP-ratio}) is easily satisfied. We note that those values of the Yukawa couplings are perfectly compatible with the inflationary constraints.

Regarding detection constraints for the DM particles, our asymmetric candidate may scatter off nuclei in underground terrestrial detectors, giving rise to direct detection signature. Due to the particularity of inert fermion DM, the interaction will be inelastic and mediated by the $Z$ boson. For this kind of scattering, a Majorana splitting of about 100 keV can explain the DAMA modulated signal and is only partially excluded by the upper limit of Xenon100~\cite{Aprile:2011ts} released in 2011 and which is the most stringent bound up to now. Therefore $\psi_{\rm DM}$ as inelastic candidate is allowed in all the range from 45 GeV up to 350 GeV. Further details about direct detection of fermion doublet DM can be read from~\cite{Arina:2011cu}.

\section{Conclusions}
\label{sec:concl}

The indication at the Large Hadron Collider (LHC) of a SM like Higgs boson with mass around 125-126 GeV suggests that the SM vacuum might be metastable at around $10^{9}$ GeV, although it can be extended up Planck scale by considering the present theoretical and experimental uncertainties. In this paper we studied the scalar triplet extension of the SM which not only evades the possibility of having a metastable vacuum at least up to the unitarity scale, $\calO(10^{14})$ GeV, of the theory but also has a rich phenomenology in presence of a vectorial fermionic doublet stabilized by means of a remnant $Z_2$ flavour symmetry and thereby playing the role of a DM candidate.

We introduced non-minimal couplings to gravity for both scalar triplet and the SM Higgs. In presence of these couplings the scalar triplet, mixed with the SM Higgs, drives inflation in the early Universe. We showed that the extended scalar potential gives rise to slow-roll single field inflation, once the heavy field is stabilized at a minimum of the potential. In general the inflaton is an admixture of triplet scalar and the SM Higgs. However, depending on the minimum, the inflaton could either be a triplet scalar or be a SM Higgs. Taking into account that the potential should be positive definite, these three scenarios give different constraints on the quartic couplings, namely $\lambda_\Delta$ and $\lambda_{H\Delta}$. We recall that the quartic coupling $\lambda_H$ is fixed assuming that the Higgs mass is around 125 GeV.

Unfortunately it is not possible to measure the quartic couplings of the triplet at the LHC, because of its large mass. Hence it is not possible to distinguish between the three type of inflationary pictures. Also the inflationary scenario does not constrain the Yukawa couplings of scalar triplet to DM and leptons. It only constrains the dimensionful coupling $\mu_H$ between scalar triplet and the SM Higgs to be smaller than $10^{-7} M_{\rm pl}$ in order not to destabilize the scalar potential. This is also in agreement with another constraint arises from the RG evolution of the quartic coupling of the SM Higgs, which shows that the scale of new physics should be order of $10^8$ GeV in order not to jeopardize the stability of the SM Higgs potential. Based on these constraints we set the mass scale of triplet to be $\simeq \calO (10^8)$ GeV such that it not only stabilized the scalar potential but also gave masses to active neutrinos via type-II seesaw.

Since the triplet couples to leptons and fermion doublet DM and in general these couplings are complex, its out-of-equilibrium decay produce asymmetries simultaneously in either sectors. The lepton asymmetry produced by the triplet decay can be converted to observed baryon asymmetry in presence of the EW sphalerons. The relic abundance of DM can be accounted by an asymmetric component rather than the symmetric component which is usually generated by the freeze-out mechanism. Since DM is a doublet its mass is necessary to be larger than 45 GeV in order not to increase the invisible $Z$ width. Moreover, DM is inelastic and therefore scattered-off nuclei through $Z$ boson. Since the mass splitting between the two companions of DM is about 100 KeV, an order of 100 GeV DM can explain the annual modulation signal at DAMA and the null result of Xenon100.

An interesting possibility will be to perform a detailed numerical treatment of the inflation scenarios taking into account the variation due to the reheating parameter and of multi-field dynamics, on the lines of~\cite{Ringeval:2007am,Martin:2010kz}. Another attractive possibility is to lower the triplet mass scale down to TeV scale, in that case same sign dilepton signal will be accessible at the LHC. The neutrino masses are preserved by means of a variant of the type-II seesaw, which involves two scalar triplets~\cite{Majee:2010ar,McDonald:2007ka}. We leave these investigations for further works.

\section*{Acknowledgements}
CA acknowledges use of the cosmo computing resource at CP3 of Louvain University.
JG is partially supported by a Korean-CERN Fellowship.

\appendix

\section{Boltzmann equations for quasi-equilibrium evolution of triplet scalars}
\label{appA}

We briefly report the relevant Boltzmann equations relating the generation of the CP asymmetries in the dark and leptonic sectors. A more  in depth discussion can be found in~\cite{Arina:2011cu} and references therein.

The evolution of $\zeta_1^\pm$ density is described the following Boltzmann equation:
\begin{equation}
\frac{dX_{\zeta}}{dz}=-\frac{\Gamma_D}{zH(z)}\left( X_{\zeta} -X_{\zeta}^{\rm eq} \right) -
\frac{\Gamma_A}{z H(z)} \left( \frac{X_{\zeta}^2-{X_{\zeta}^{\rm eq}}^2}{X_{\zeta}^{\rm eq}} \right)\,,
\label{boltzman-1}
\end{equation}
with $z=M_1/T$ and $X_\zeta \equiv n_{\zeta_1^-}/s = n_{\zeta_1^+}/s$, if the mass of the triplet stays the same after EW symmetry breaking. The decay term is described by
\begin{equation}
\Gamma_D=\Gamma_1 \frac{K_1(z)}{K_2(z)}\,,
\end{equation}
while the scattering term and the scattering densities are given by:
\begin{align}
\Gamma_A = & \, \, \frac{\gamma_A}{n_{\zeta_1}^{\rm eq}}\,,
\\
\gamma ( \zeta_1^+\zeta_1^-\to \bar{f}f)  =& \, \, \frac{M_1^4 \left(6 g_2^4 + 5 g_Y^4\right)}{128 \pi^5 z} \int_{x_{\rm min}}^{\infty} dx \sqrt{x} K_1(z \sqrt{x}) r^3 \,,
\\
\gamma ( \zeta_1^+\zeta_1^-\to H^\dagger H) =& \, \, \frac{M_1^4 \left(g_2^4 + g_Y^4/2\right)}{512\pi^5 z} \int_{x_{\rm min}}^{\infty} dx \sqrt{x} K_1(z \sqrt{x})r^3 \,,
\\
\gamma ( \zeta_1^+\zeta_1^-\to W^a W^b) =&\,\,  \frac{ M_1^4 g_2^4}{64 \pi^5 z} \int_{x_{\rm min}}^{\infty} dx \sqrt{x} K_1(z \sqrt{x})
\left[r \left(5+\frac{34}{x}\right)-\frac{24}{x^2}(x-1)\log \left(\frac{1+r}{1-r} \right)\right]\,,
\\
\gamma ( \zeta_1^+\zeta_1^-\to BB) =& \,\, \frac{3 M_1^4 g_Y^4}{128 \pi^5 z}
\int_{x_{\rm min}}^{\infty} dx \sqrt{x} K_1(z \sqrt{x}) \left[r\left(1+\frac{4}{x}\right)-\frac{4}{x^2}(x-2) \log \left( \frac{1+r}{1-r} \right) \right]\,,
\\
\gamma ( \zeta_1^+\zeta_1^-\to \bar{\psi}\psi) = &\,\, \frac{M_1^4 \left(6 g_2^4 + 5 g_Y^4\right)}{128 \pi^5 z} \int_{x_{\rm min}}^{\infty} dx \sqrt{x} K_1(z \sqrt{x}) r^3 \,,
\end{align}
where $H(z)= H(T=M_1)/z^2$, $r= \sqrt{1-4/x}$ and $x=\hat{s}/M_1^2$.

The asymmetry $Y_{\zeta} = ( n_{\zeta_1^-}-n_{\zeta_1^+})/s$ evolves due to the decay and inverse decay of $\zeta_1^\pm$ particles. The corresponding Boltzmann equation is given by
\begin{equation}
\frac{d Y_{\zeta}}{dz} = -\frac{\Gamma_D}{zH(z)} Y_{\zeta} + \sum_j \frac{\Gamma^j_{ID}}{zH(z)}
2 B_j Y_j\,,
\label{boltzman-2}
\end{equation}
where $Y_j=(n_j-n_{\bar j})/s$, with $j=L, H, \psi$ and
\begin{equation}
\Gamma^j_{ID} = \Gamma_D \frac{X_{\zeta}^{\rm eq}}{X_j^{\rm eq}} \quad {\rm and} \quad B_j=\frac{\Gamma_j}
{\Gamma_1}\,,
\label{eq:pippo}
\end{equation}
where $X_j=n_j/s$. The evolution of the asymmetries $Y_j$ is given by the Boltzmann equation:
\begin{equation}
\frac{d Y_j}{dz} =\ 2\  \left\{ \frac{\Gamma_D}{zH(z)} \left[ \epsilon_j (X_{\zeta} - X_{\zeta}^{\rm eq}) \right] +  B_j \left( \frac{\Gamma_D}{zH(z)} Y_{\zeta}  - \frac{\Gamma^j_{ID}}{zH(z)} 2 Y_j\right) -\sum_k \frac{\Gamma^k_S}{z H(z)} \frac{X_{\zeta}^{\rm eq}}{X_k^{\rm eq}} 2 Y_k\right\}\,,
\label{boltzman-3}
\end{equation}
where $\Gamma_S=\gamma_S/n_{\zeta_1}^{\rm eq}$ is the scattering rate involving the number violating processes, such as $LL \to \zeta_1 \to HH$. The front factor in (\ref{boltzman-3}) takes into account of the two similar particles produced in each decay. Note that because of the conservation of hypercharge the Boltzmann equations~(\ref{boltzman-1}), (\ref{boltzman-2}) and~(\ref{boltzman-3}) satisfy the relation: $2 Y_\zeta + \sum_j Y_j =0$.

\bibliographystyle{elsarticle-num.bst}
\bibliography{biblio}

\end{document}